\documentclass[]{aa}
\usepackage{txfonts}
\usepackage{natbib}
\usepackage{graphicx}
\usepackage{psfig}
\usepackage{changes}

\def\nustar{{\em NuSTAR}}
\def\src{IGR~J11215--5952}
\def\hd{HD~306414}

\def\cyclabs{CRSFs}
\def\cyclab{CRSF}

\def\xmm{{\em XMM--Newton}}
\def\sw{{\em Swift}}
\def\inte{{\em INTEGRAL}}

\def\gaia{{\em Gaia}}

\def\m_one{{\em MOS1}}
\def\m_two{{\em MOS2}}

\def\approxgt{\mathrel{\hbox{\rlap{\lower.55ex \hbox {$\sim$}}
        \kern-.3em \raise.4ex \hbox{$>$}}}}
\def\approxlt{\mathrel{\hbox{\rlap{\lower.55ex \hbox {$\sim$}}
        \kern-.3em \raise.4ex \hbox{$<$}}}}
\def\ps {$P_{\rm spin}$}

\def\flux {\mbox{erg cm$^{-2}$ s$^{-1}$}}

\def\ltsima{$\; \buildrel < \over \sim \;$}
\def\lsim{\lower.5ex\hbox{\ltsima}}
\def\gtsima{$\; \buildrel > \over \sim \;$}
\def\gsim{\lower.5ex\hbox{\gtsima}}

\def\ergsec{\hbox{erg s$^{-1}$}}

\def\hcm {\hbox {\ifmmode $ atom cm$^{-2}\else atom cm$^{-2}$\fi}}

\def\arcsec {\hbox{$^{\prime\prime}$}}

\def \ATel {Astron.\ Tel.}

\def \apj {ApJ}
\def \aj {AJ}
\def \apjl {ApJL}
\def \apjs {ApJS}
\def \aap {A\&A}

\def \mnras {MNRAS}

\def \ssr {Space Science Reviews}

\newcommand{\be}{\begin{equation}}

\newcommand{\ee}{\end{equation}}

\begin{document}
\title{\nustar\ observation of the Supergiant Fast X-ray Transient \src\ during its 2017 outburst}
\author{L.~Sidoli\inst{1}, K.~Postnov\inst{2,3}, A.~Tiengo\inst{4,5,1}, P.~Esposito\inst{4,1}, V.~Sguera\inst{6}, A.~Paizis\inst{1}, and G.A.~Rodr\'iguez~Castillo\inst{7}
}
\institute{$^1$ INAF, Istituto di Astrofisica Spaziale e Fisica Cosmica, via A.\ Corti 12, I-20133 Milano, Italy
\\
$^2$ Sternberg Astronomical Institute, M.V. Lomonosov Moscow State University, 13 Universitetskij pr., 119234 Moscow, Russia\\
$^3$ Kazan Federal University, Kremlevskaya 18, 420008 Kazan, Russia
\\
$^4$ Scuola Universitaria Superiore IUSS Pavia, Palazzo del Broletto, piazza della Vittoria 15, I-27100 Pavia, Italy
\\
$^5$ INFN, Sezione di Pavia, Via A. Bassi 6, 27100 Pavia, Italy 
\\
$^6$ INAF, Osservatorio di Astrofisica e Scienza dello Spazio, Area della Ricerca del CNR, via Gobetti 101, I-40129 Bologna, Italy
\\
$^7$ INAF, Osservatorio Astronomico di Roma, via Frascati 33, I-00040 Monteporzio Catone (Roma), Italy
}

\offprints{L. Sidoli, lara.sidoli@inaf.it}

\date{Received 3 April 2020 / Accepted 17 April 2020}

\authorrunning{L. Sidoli et al.}

\titlerunning{\nustar\ observes \src\ in outburst in 2017}

\abstract{
We report on the results of a \nustar\ observation of the Supergiant Fast X-ray Transient pulsar \src\ during the peak of its outburst in June 2017. 
\src\ is the only SFXT undergoing strictly periodic outbursts, every 165~days. 
\nustar\ caught several X-ray flares, spanning a dynamic range of 100, and detected X-ray pulsations at 187.0~s, consistent with previous measurements.
The spectrum from the whole observation is well described by an absorbed power-law  (with a photon index of 1.4) modified, above $\sim$7 keV, by a cutoff with an e-folding energy of $\sim$24~keV. %
A weak emission line is present at 6.4~keV, consistent with K$_{\alpha}$ emission from cold iron in the supergiant wind. 
The time-averaged flux is $\sim$1.5$\times10^{-10}\flux$ (3-78 keV, corrected for the absorption),
translating into an average luminosity of about 9$\times$10$^{35}$~\ergsec (1--100 keV, assuming a distance of 6.5 kpc). 
The \nustar\  observation allowed us to perform the most sensitive search for  cyclotron resonant scattering features in the hard X-ray spectrum, resulting in no significant detection in any of the different spectral extractions adopted (time-averaged, temporally-selected, spin-phase-resolved and intensity-selected spectra).
The pulse profile showed an evolution with both the energy (3-12 keV energy range compared with 12-78 keV band) and the X-ray flux: a double peaked profile was evident at higher fluxes (and in both energy bands), while  a single peaked, sinusoidal profile was present at the lowest intensity state achieved within the \nustar\ observations (in both energy bands). 
The intensity-selected analysis allowed us to  observe an anti-correlation of the pulsed fraction with the X-ray luminosity.
The pulse profile evolution can be explained by X-ray photon 
scattering in the accreting matter above 
magnetic poles of a neutron star at the quasi-spherical settling accretion stage.
}

\maketitle

        \section{Introduction\label{intro}}

Supergiant Fast X-ray Transients (SFXTs) are a sub-class of High Mass X-ray Binaries (HMXBs;  \citealt{Walter2015, Martinez-Nunez2017, Sidoli2017review, Sidoli2018}) which show sporadic, short, and recurrent X-ray transient emission \citep{Sguera2005, Sguera2006} and are associated with early-type supergiants \citep{Negueruela2006}.  
At odds with classical HMXBs hosting a blue supergiant plus a neutron star (NS, hereafter), systems known since the early times of X-ray astronomy \citep{White1983}, SFXTs are not persistent X-ray emitters. 
SFXT  outbursts span a few days, with a peak usually lasting  less than one day, and are punctuated by short X-ray flares whose duration  distribution   peaks in the range 100-2000 seconds \citep{Sidoli2019}. 
Their X-ray spectrum is similar to that of classical accreting pulsars, and X-ray pulsations have been discovered in a few members of the class (ranging from $\sim$20 to 1200 seconds, \citealt{Sidoli2017review}), a signature for the 
presence of a NS. 
There is no consensus in the literature about the mechanism powering the transient X-ray emission, also because some of the crucial NS properties needed to investigate the transition across different regimes, from quiescence to outburst \citep{Bozzo2008, Shakura2012, Shakura2017}, are elusive in SFXTs: it is worth noting that there is no SFXT, to date, where both the pulsar spin period and its magnetic field are known with certainty.

\src\ is an X-ray transient source discovered in 2005 \citep{Lubinski2005} with the \inte\ satellite \citep{winkler2003}, later 
associated with the B0.5\,Ia star \hd\ \citep{Negueruela2005b}. 
The presence of a supergiant companion together with the short outbursts (a few thousand seconds, as observed by \inte, \citealt{SidoliPM2006}) make it a member of the SFXT sub-class.
\src\ is also an X-ray pulsar (spin period, \ps, of 186.78$\pm{0.3}$~s, discovered by \citealt{Swank2007}) and undergoes an X-ray outburst every $\sim$165~days \citep{Sidoli2007, Romano2009}, a remarkable unique property among SFXTs 
(see \citealt{Sidoli2017review} for the most recent review focussed on SFXTs).
 
The periodicity in the occurrence of the outburst is assumed to be due to the orbital period of the system \citep{SidoliPM2006}, 
making it the supergiant HMXB with the longest orbital period, overlapping with the region occupied by  Be/X-ray binary systems in the so-called Corbet diagram \citep{Corbet1986}.
An extreme orbital eccentricity ($e$$\ga0.8\:$) is suggested by the shortness of its outburst compared with the  long orbital period \citep{Romano2007, Sidoli2007, Romano2009, Negueruela2008, Lorenzo2014}. The extreme orbital geometry is not contradicted by the observed radial velocity curve of the companion \hd, which is  dominated by other effects than orbital ones, most likely stellar pulsations \citep{Lorenzo2014}.

The observed properties of the X-ray outburst led \citet{Sidoli2007} to propose that the supergiant wind is magnetically compressed along the stellar equatorial plane (similar to the mechanism discussed by \citealt{Ud-Doula2002}), in order to produce a short duration outburst when the NS crosses this denser wind component along the orbit. This hypothesis is supported by the  detection of a stellar magnetic field of $\sim$1~kG in the optical companion \citep{Hubrig2018}. The possible role played by the supergiant magnetic field in triggering SFXT outbursts has been proposed by \citet{Shakura2014} in the quasi-spherical accretion scenario.

The parallax of the star \hd\ has been measured by \gaia\ (Data Release 2, DR2, \citealt{GaiaDR2}), translating into a distance of 6.5$^{+1.4} _{-1.0}$~kpc \citep{Bailer-Jones2018}. 
The source distance previously reported in the literature was very close to this value, 
from $d = 6.2$ kpc \citep{Masetti2006} to 
$d \gtrsim 7$ kpc \citep{Lorenzo2014}.
\hd\ was observed at optical and UV wavelengths, with controversial  results about the stellar  properties (in particular, the wind mass loss rate and the wind velocity), due to different assumptions made
\citep{Lorenzo2014, Hainich2020}.

A \nustar\ observation  of \src\ that was performed during an outburst occurred in 2016 (with a net exposure of 20 ks), led to a hint (at 2.63$\sigma$) of the presence of  a cyclotron resonant scattering feature (CRSF, hereafter) at 17 keV \citep{Sidoli2017}. Being produced by scattering of X--rays by electrons in quantized energy levels in the strong magnetic field of the pulsar, CRSFs are the only direct way to measure the NS magnetic field strength. In \src\ the suggested feature would imply a NS magnetic field of 
2$\times10^{12}$~G \citep{Sidoli2017}.
These findings make the pulsar \src\ the perfect target for a more in-depth 
investigation of its hard X-ray spectrum with \nustar, searching for a confirmation of the \cyclab.

         \section{Observation and data reduction}
         \label{data}

\subsection{\nustar}

The Nuclear Spectroscopic Telescope Array (\nustar, \citealt{Harrison2013}) carries two co-aligned telescopes which focus X-ray photons onto two independent
Focal Plane Modules named A and B (hereafter FPMA and FPMB),
with a 12$'$$\times$12$'$ field-of-view (FOV).
Each FPM contains four (2$\times$2) solid-state cadmium zinc telluride (CdZnTe) pixel detectors
providing a spatial resolution of 18$''$ (full width at half maximum) and a spectral resolution of 
400~eV (FWHM) at 10 keV, with an effective area calibrated in the energy band 3-78~keV \citep{Madsen2015}.

The \nustar\ observation of the source \src\ covered an entire day,
starting on  2017 June 21 at 08:58 and ending  on 2017 June 22 at 08:31 (TT). 
It was a fixed time observation, planned to cover the brightest phase of the outburst,
as expected from the known strict periodicity in the outburst recurrence of 164.6~d \citep{Romano2009}. 
The expectations were  confirmed by a monitoring with \sw/XRT  around the expected peak 
(see Sect.~\ref{sect:sw} and Fig.~\ref{fig:swnulc}).
The low satellite orbit (with data-gaps lasting about 30~minutes 
every revolution) reduced the net exposure time to 43.7 ks.

\nustar\ data (Obs.ID 30301010002) were processed using the version 1.7.1 
of the \nustar\ data analysis software ($NuSTAR~DAS$).
Spectra and light curves were extracted with {\em nuproducts} 
on the cleaned event files, using circular extraction regions with a radius of 60\arcsec.
The background spectrum was extracted away from the point-spread function (PSF) source wings.
Since the background was stable and constant along the observation, no further filtering was applied.
\nustar\ source light curves with arrival times corrected to the Solar System barycenter have been extracted
with the $NuSTAR~DAS$ tool {\em nuproducts} and the keyword ``barycorr=yes''.
When needed (e.g., in  pulse phase spectroscopy), good-time intervals (GTIs)  were generated using {\em xselect} and then running 
{\em nuproducts} using the ``usrgtifile'' keyword, to correctly extract the temporal selected spectra.
The  source net count rate in the energy range 3-78 keV from the overall exposure (43.7~ks) was
$\sim$1.5 counts~s$^{-1}$ per single module.

Spectra from FPMA and FPMB in the energy band 3-78 keV 
were simultaneously fitted in {\sc xspec} (v.12.10.1;  \citealt{Arnaud1996}), 
adopting cross-calibration constants to take into account calibration uncertainties. 
All fluxes reported in this paper have been estimated using 
the FPMA response matrix as reference.
The same procedure was used when \nustar\ spectra were jointly fitted together with the \sw/XRT one.

When fitting the spectra in {\sc xspec}, the absorption model {\sc TBabs} was adopted, assuming the photoelectric absorption cross sections of \cite{verner1996} and the interstellar abundances of \cite{wilms2000}. The spectra were re-binned to have at least 20 counts per bin, to apply the $\chi^{2}$ statistics. 
All  uncertainties  in the spectral analysis are given at 90\% confidence level, for one interesting parameter.

All \nustar\ light curves and pulse profiles reported in this paper are background subtracted.

\subsection{\sw}
\label{sect:sw}

The Neil Gehrels \sw\ observatory (\citealt{Gehrels2004}; \sw\ hereafter) monitored the source X-ray flux (0.3-10 keV) during the expected times of the 2017 outburst. 
We reprocessed the \sw/XRT observations (X-ray Telescope XRT; \citealt{Burrows2005:XRTmn}) 
using {\em xrtpipeline}   in the  {\em HEASoft }   software package  version  6.25, adopting standard procedures.  
The source flux was low and only photon-counting data (PC) were used.

The appropriate spectral redistribution matrices were used, available in the Calibration Database maintained by the High Energy Astrophysics Science Archive Research Center (HEASARC). Source events were extracted from a circular region with a radius of 20 pixels ($\sim$47\arcsec), while local background events were taken from an annular region, centered on the source, with inner and outer radii of 30 and 60 pixels, respectively. 

The \sw/XRT spectrum was analyzed in the energy range 2-8 keV, since there are no net counts both below 2 keV and above 8 keV.

\section{Temporal analysis}
\label{timing}
  
\subsection{Light curves}  
 
\begin{figure}
\begin{center}
\includegraphics[width=5.75cm,angle=-90]{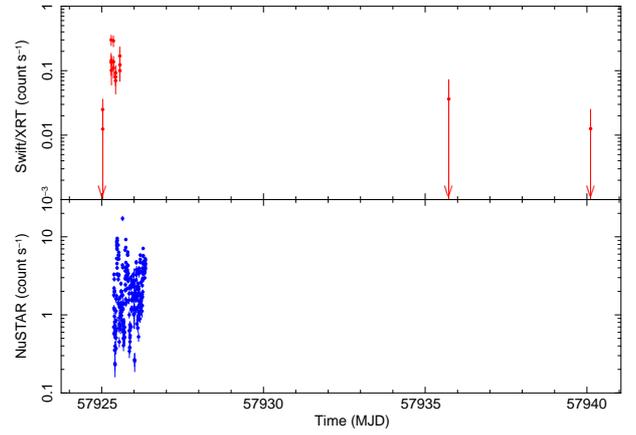}
\caption{\sw/XRT (0.3-10 keV) light curve of \src\ to monitor the onset of the 2017 outburst (upper panel), compared with the times of the \nustar\ observation (lower panel, 3-78 keV). 
A time binning of 187 s has been assumed. We note that, since there are gaps in both observations (by \sw\ and \nustar), unfortunately they were never simultaneous. 
The down arrows in the upper panel indicate upper limits.
}
\label{fig:swnulc}
\end{center}
\end{figure}

The source light curve during the 2017 outburst is shown in Fig.~\ref{fig:swnulc}, observed with  \sw/XRT and  \nustar\  
 (in the upper and lower panel, respectively). The \sw\ monitoring around the times of the \nustar\ exposure, although with a
large gap of about ten days, due to Sun constraints which prevented additional \sw/XRT snapshots, indicates that the \nustar\ pointing caught the source at the peak of the outburst.
 
\begin{figure}
\begin{center}
\includegraphics[width=6.15cm,angle=-90]{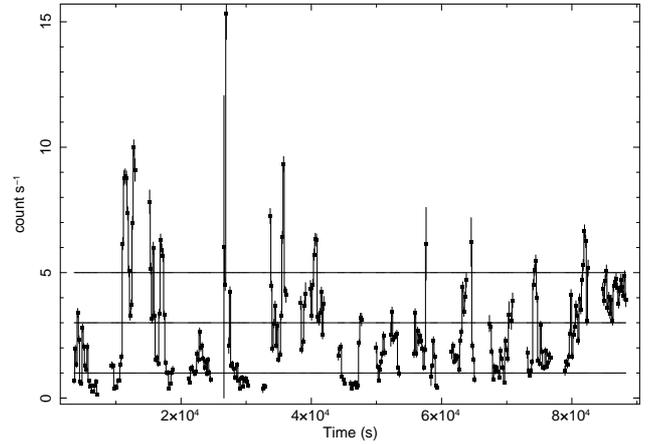}
\caption{\nustar\ net source light curve (3--78 keV) of the 2017 outburst. 
A time binning of 187 s has been assumed. The three horizontal lines mark the intensity intervals adopted in the intensity selected spectroscopy (see Sect.~\ref{sec:intsel} for details).  
}
\label{fig:nulc}
\end{center}
\end{figure}

Fig.~\ref{fig:nulc} reports the \nustar\ net source light curve alone, 
adopting a time bin of 187~s, the known pulsar spin period, in order to avoid intensity
variability due to the X-ray pulsations. The three horizontal lines divide the outburst into four intensity regions, which have been 
adopted to perform the intensity-selected spectroscopy (Sect.~\ref{sec:intsel}).

\begin{figure*}
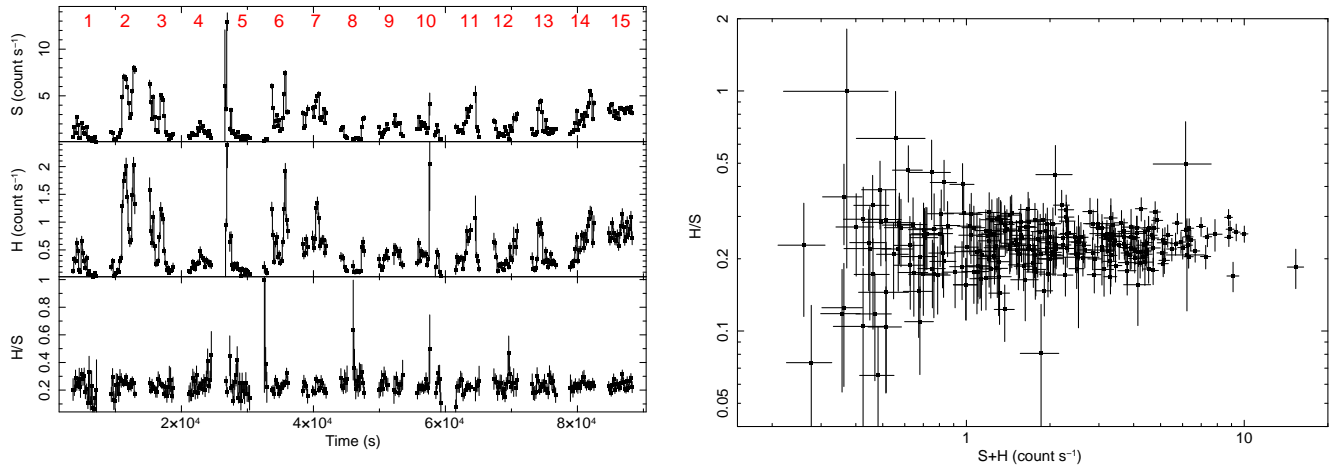

\begin{center}
\includegraphics[width=5.9cm,angle=-90]{fig03a.ps} 
\includegraphics[width=6.1cm,angle=-90]{fig03b.ps} \\
\caption{\src\ light curves in two energy bands (S=3--12 keV, H=12--78 keV) together with their hardness ratio (HR, on the left). Red numbers on the top of the soft X-rays light curve indicate the \nustar\ satellite revolutions analyzed in Table~\ref{tab:revsel}. On the right, the same hardness ratio versus the 3--78 keV count rate is shown. In both panels, a bin time of 187~s has been adopted.
}
\label{fig:nulchr_int}
\end{center}
\end{figure*}

In Fig.~\ref{fig:nulchr_int} we show the net light curves extracted from two energy bands, below (S) and above 12 keV (H),
together with their hardness ratio (H/S). These two energy ranges were chosen to enable a more direct
comparison with previous observations \citep{Sidoli2017}. 
There is no evidence of a strong trend of the spectral hardness, both with time (Fig.~\ref{fig:nulchr_int}, left panel) 
 and with the source intensity (right panel). 
 The huge range of variability of the X-ray flux, spanning a factor of 100 (from 0.15 to 15 count s$^{-1}$, in the range 3-78 keV) is also evident.

\subsection{Pulse profiles}  
 
\begin{figure}
\includegraphics[scale=0.36,angle=-90]{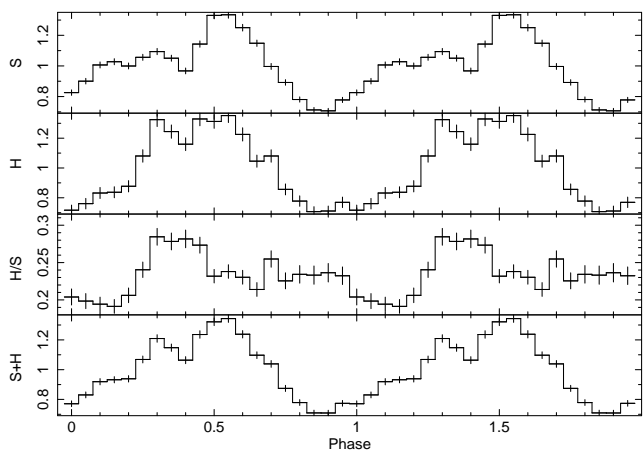} \\ 
\caption{\src\ pulse profile folded on the spin period of 187~s, assuming MJD 57432 as epoch of phase zero. The whole \nustar\ exposure time has been considered.
From top to bottom we show the soft profile (S=3-12 keV), the hard one (H=12-78 keV),
their hardness ratio (H/S) and the profile in the total band (S+H = 3-78 keV), respectively.
Each pulse profile has been normalized by dividing by the average source intensity in the considered energy band.
}
\label{fig:pulseprofile_entire}
\end{figure}

We performed the timing analysis on \nustar\ background subtracted light curves (3-78 keV, bin time of 0.1 s), after correcting the arrival times to the Solar System barycenter. Using epoch folding techniques, we searched for the known rotational period, measuring a periodicity \ps=187.0$\pm{0.12}$ s (1$\sigma$), consistent with previous values \citep{Swank2007, Sidoli2007, Sidoli2017}.
Fig.~\ref{fig:pulseprofile_entire} shows the pulse profile obtained folding the whole exposure time on this periodicity,
in three energy ranges (3-12 keV, 12-78 keV, 3-78 keV), together with the hardness ratio of 12-78 keV to 3-12 keV (third panel from the top). Variability of the spectral hardness is evident along the rotation of the pulsar.

We calculated the pulsed fraction, $PF$ (defined as $PF\equiv(F_{max}-F_{min})/(F_{max}+F_{min}$),
where $F_{max}$ and $F_{min}$ are the count rates at the maximum and at the minimum of the spin profile, respectively), in three energy ranges, finding a correlation with the energy: $PF$=$30\pm{2}$~\% (3-12 keV), 
$PF$=$35\pm{2}$~\% (12-20 keV) and 
$PF$=$43\pm{4}$~\% (20-78 keV). 
The pulsed fraction measured over the entire \nustar\ energy range is $PF$=$30.1\pm{0.9}$~\% (3-78 keV).
 
\begin{figure}
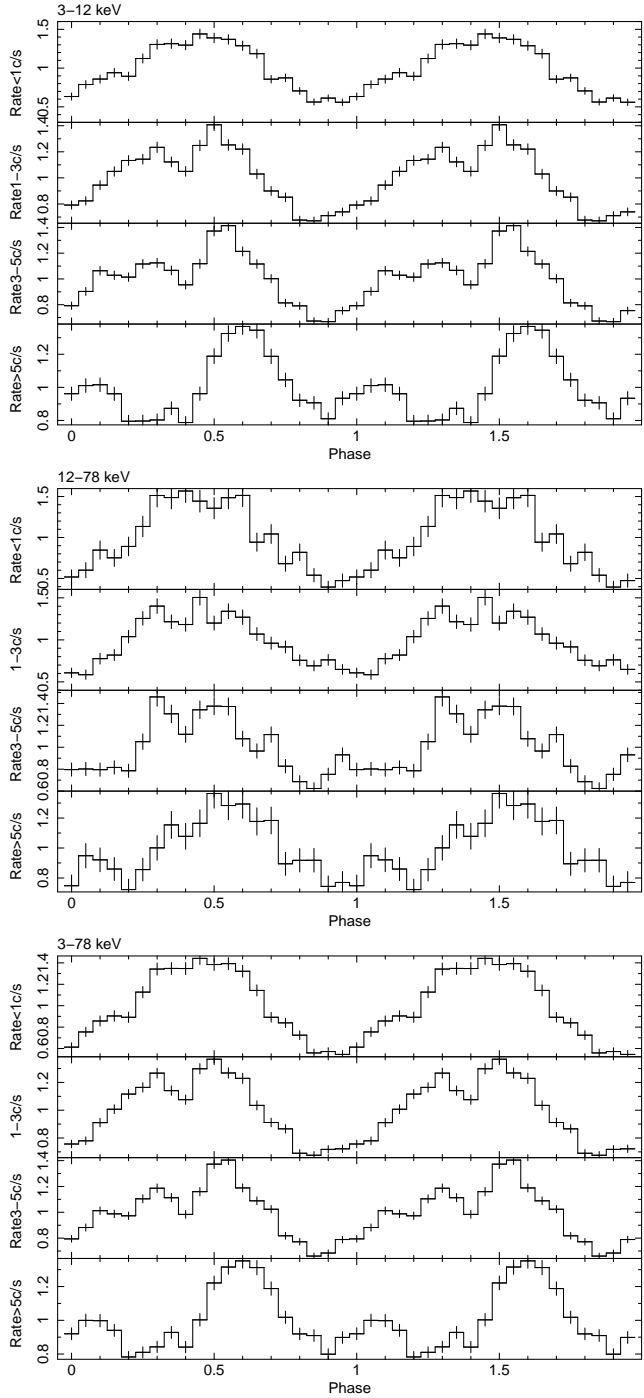

\includegraphics[scale=0.36,angle=-90]{fig04a.ps} \\
\includegraphics[scale=0.36,angle=-90]{fig04b.ps} \\
\includegraphics[scale=0.36,angle=-90]{fig04c.ps} \\
\caption{\src\ pulse profiles in three energy bands (from top to bottom, 3-12 keV, 12-78 keV and 3-78 keV). For each energy band, we report pulse profiles in four intensity-selected intervals. 
Each profile has been normalized by dividing by the average source intensity in the considered energy band.
}
\label{fig:pulse_b}
\end{figure}

\begin{figure}
\begin{center}
\hspace{-1.1cm}
\includegraphics[width=9.5cm,angle=0]{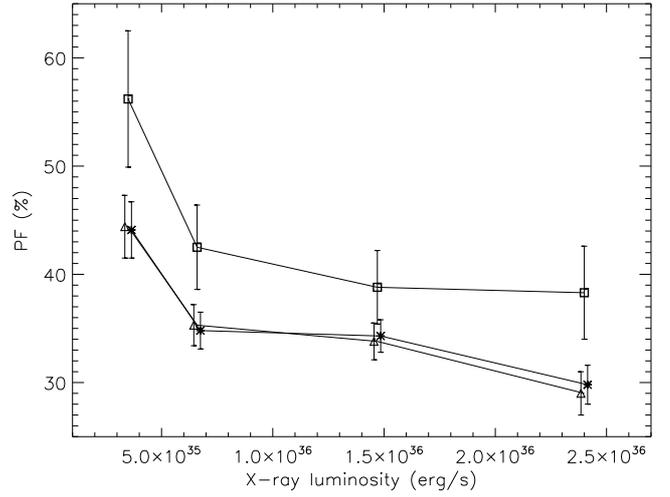}
\caption{Pulsed fraction dependence on the average X-ray luminosity (1-100 keV) measured from the four intensity selected spectra (Table~\ref{tab:intsel}). Results in three energy bands are shown with the following symbols: triangles mark the pulsed fractions in the energy band 3-12 keV, squares in the band 12-78 keV, stars in the total band 3-78 keV. A small shift on the x-axis has been applied for visualization purposes.
}
\label{fig:pf}
\end{center}
\end{figure}

We investigated the evolution of the pulse shape with the X-ray luminosity, extracting the folded light curves in the four intervals of  count rate shown in Fig.~\ref{fig:nulc} and used for the intensity selected spectroscopy (Table~\ref{tab:intsel}): 
count rate interval $<$1~count~s$^{-1}$, 
1-3~count~s$^{-1}$, 
3-5~count~s$^{-1}$ and 
count rate  $>$5~count~s$^{-1}$ (3-78 keV, time bin of 187 s). 
These intensity ranges correspond to the average X-ray luminosities reported in the seventh column of Table~\ref{tab:intsel}. 
In Fig~\ref{fig:pulse_b} we have reported the intensity selected pulse profiles in three energy bands. 
The pulse profile changes also with the X-ray luminosity, and an anti-correlation of the pulsed fraction 
with the X-ray luminosity is observed (Fig.~\ref{fig:pf}).

        \section{Spectroscopy}
        \label{sec:spec}

We have performed spectroscopy extracting  \nustar\ spectra in different ways (time-averaged, temporally-resolved,  intensity-selected and spin-phase-resolved spectra), to search for evidence of \cyclabs, as reported in the following sub-sections.

\subsection{Time-averaged spectrum}
\label{sec:av_spec}

We first analyzed the \nustar\ spectra (FPMA and FPMB) 
extracted from the entire exposure time, fitting them jointly together
with the \sw/XRT spectrum (but we note that XRT snapshots were never simultaneous with \nustar).

The \sw/XRT exposure time is 1.8~ks and the source 
average count rate in the 0.3-10 keV energy band is 0.117 $\pm{0.008}$~count~s$^{-1}$.

A single power-law is not able to  model the spectrum appropriately, given the presence of
a cutoff at high energies. A faint emission line is evident at $\sim$6.4\,keV (equivalent width, EW$\sim$50\,eV)
consistent with being produced by neutral iron, modeled with a narrow Gaussian line. This emission line is significantly detected
only in the time-averaged spectrum.

\begin{figure}
\begin{center}
\includegraphics[width=5.8cm,angle=-90]{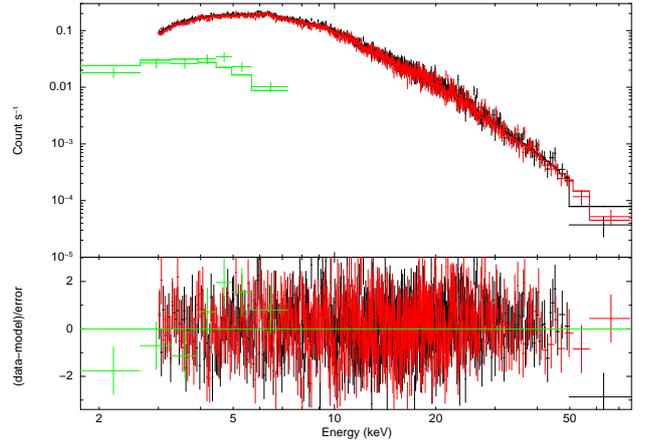}
\caption{\sw/XRT and \nustar\ time-averaged spectra, fitted with an absorbed power-law modified by a high energy cutoff, together with a Gaussian line in emission, accounting for iron line at $\sim$6.4~keV (see Table~\ref{tab:av_spec}, model PLCUT, for the best fit parameters). In the lower panel, the residuals are shown, in terms of standard deviation.}
\label{fig:av_spec}
\end{center}
\vspace{-0.75 cm}
\end{figure}

In Table~\ref{tab:av_spec} we report the results with two models: 
the first one (which we name PLCUT, hereafter) includes a power-law modified by a high energy cutoff ({\sc highecut} in {\sc xspec}), 
this latter 
defined as M(E)=${ (exp[(E_{cut}-E)/E_{fold}] )}$ when E$ \ge E_{cut}\:$, while M(E)=1 at E$ \le E_{cut}\:$; 
the second model (named CUTOFFPL) adopts a cutoff power-law ({\sc cutoffpl} in {\sc xspec}), 
defined as  
A(E)=KE$^{-\Gamma}exp[(-E/E_{cut})]$, where $\Gamma$ is the photon index.
The model 
PLCUT was successfully adopted by \citet{Sidoli2017} to investigate the 2016 \nustar\ observation, as a statistically acceptable description.
Since the best-fit has been obtained with PLCUT (Fig.~\ref{fig:av_spec}),  in the following we will report results adopting this spectral 
model only. We show the count spectra fitted with this model in Fig.~\ref{fig:av_spec}. 
We found no evidence of statistically significant absorption features.
 
We note that the analysis of the \nustar\ spectra alone led to spectral
parameters consistent with the ones reported in Table~\ref{tab:av_spec}.

%
\begin{table}
\caption[]{Broad-band (\sw/XRT together with \nustar) time-averaged spectrum (0.3--78 keV; see Sect.~\ref{sec:av_spec}). 
}
\begin{tabular}{lcc}
 \hline
\noalign {\smallskip}
Parameter                                       &  Model PLCUT                  & Model CUTOFFPL         \\
\hline
\noalign {\smallskip}
N$_{\rm H}$ ($10^{22}$~cm$^{-2}$)               &  $5.28 ^{+0.72} _{-0.76}$  &  $6.28 ^{+0.65} _{-0.65}$ \\
$\Gamma^a$                                      &  $1.46 ^{+0.05} _{-0.07}$  &  $1.33 ^{+0.05} _{-0.05}$  \\
E$_{cut}$ (keV)                                 &  $7.24 ^{+0.46} _{-0.66}$  &  $20.8 ^{+1.5} _{-1.4}$    \\
E$_{fold}$ (keV)                                &  $23.6 ^{+1.9} _{-1.7}$    &   $-$  \\
\hline
E$_{line}$  (keV)                               &  $6.32 ^{+0.08} _{-0.03}$  &   $6.36 ^{+0.04} _{-0.09}$    \\
$\sigma_{line}$ (keV)                           &  $<0.8$                    &   $<0.8$        \\
Flux$_{line}$ (10$^{-5}$~ph~cm$^{-2}$~s$^{-1}$) &  4.2$\pm{0.1}$             &   4.5$\pm{0.1}$  \\
EW$_{line}^b$  (eV)                             &  $53\pm{13}$               &   $57\pm{13}$    \\
\hline
F$_{3-78~keV}$  (erg~cm$^{-2}$~s$^{-1}$)$^c$    &  1.39$\times$10$^{-10}$     &   1.37$\times$10$^{-10}$    \\
UF$_{3-78~keV}$    (erg~cm$^{-2}$~s$^{-1}$)$^d$ &    1.47$\times$10$^{-10}$   &    1.47$\times$10$^{-10}$     \\
UF$_{1-100~keV}$   (erg~cm$^{-2}$~s$^{-1}$)$^d$ &    1.77$\times$10$^{-10}$   &    1.78$\times$10$^{-10}$   \\
\hline
L$_{1-100~keV}$    (erg~s$^{-1}$)$^e$   &   ($8.9 ^{+4.3} _{-2.7}$)$\times$10$^{35}$ &  ($9.0 ^{+4.3} _{-2.8}$)$\times$10$^{35}$ \\
\hline
$\chi^{2}_{\nu}$/dof                    &   1.045/1131                               &     1.077/1132      \\
\hline
\label{tab:av_spec}
\end{tabular}
\footnotesize{\\
$^a$ Power-law photon index. \\
$^b$ Equivalent width. \\ 
$^c$ Flux not corrected for the absorption. \\ 
$^d$ Fluxes corrected for the absorption. \\
$^e$ The luminosity is calculated assuming a source distance of 6.5~$ ^{+1.4} _{-1.1}$~kpc. The quoted error on the luminosity has been calculated from the uncertainty on the source distance, only.}
\end{table}

\subsection{Temporally-resolved spectroscopy}
\label{sec:revsel}
 
In Table~\ref{tab:revsel} we report on the temporally-selected spectroscopy, adopting the natural segmentation 
of the \nustar\ observation into 15 satellite orbits, as evident from Fig.~\ref{fig:nulc} (marked with red numbers).
We did not find any evidence of  evolution in the spectral parameters with time, nor with the X-ray flux, within the 
uncertainties. No evidence of \cyclabs\ is present.

\begin{table*}
\caption{Temporally-resolved spectroscopy (see Sect.~\ref{sec:revsel}).}
\label{tab:revsel}
\vspace{0.0 cm}
\begin{center}
\begin{tabular}{lcccccc} \hline
 \hline
\noalign {\smallskip}
Rev.                             &   N$_{\rm H}$            &   $\Gamma^a$            &   E$_{cut}$           &   E$_{fold}$      &    UF$_{1-100~keV}^b$         & $\chi^{2}_{\nu}$/dof  \\   
                                 & ($10^{22}$~cm$^{-2}$)    &                         &   (keV)               &   (keV)           &   (erg~cm$^{-2}$~s$^{-1}$)    &                      \\     
\hline
\noalign {\smallskip}
1                                  &  $7.3^{+5.0} _{-4.4}$   & $1.40^{+0.33}_{-0.27}$  & $6.6^{+2.9}_{-6.6}$  & $22  ^{+18} _{-7}$ &       0.82$\times10^{-10}$   &    0.891/195      \\
2                                  &  $6.1^{+2.2} _{-2.7}$   & $1.34^{+0.14}_{-0.21}$  & $7.2^{+1.1}_{-1.3}$  & $22 ^{+5} _{-5}$   &       2.89$\times10^{-10}$   &    0.873/545      \\
3                                  &  $5.2^{+2.6} _{-2.6}$   & $1.51^{+0.15}_{-0.18}$  & $7.0^{+0.9}_{-1.0}$  & $26 ^{+9} _{-6}$   &       1.97$\times10^{-10}$   &    0.975/418      \\                                
4                                  &  $6.4^{+4.1} _{-4.0}$   & $1.38^{+0.49}_{-0.29}$  & $6.1 ^{+11} _{-6}$   & $25 ^{+24} _{-8}$  &       0.95$\times10^{-10}$   &    0.734/216      \\
5                                  &  $6.6^{+3.9} _{-6.6}$   & $1.51^{+0.20}_{-0.61}$  & $9.7^{+2.3}_{-4.8}$  & $24 ^{+11} _{-6}$  &       0.91$\times10^{-10}$   &    0.981/195      \\
6                                  &  $8.7^{+3.9} _{-2.5}$   & $1.62^{+0.33}_{-0.19}$  & $6.5^{+5.7}_{-6.5}$  & $36 ^{+41} _{-11}$ &       3.13$\times10^{-10}$   &    0.944/407      \\
7                                  &  $6.2^{+2.6} _{-2.5}$   & $1.54^{+0.15}_{-0.16}$  & $6.3^{+1.2}_{-0.8}$  & $28 ^{+11} _{-6}$  &       2.69$\times10^{-10}$   &    1.055/429      \\
8                                  &  $2.6^{+5.5} _{-2.6}$   & $1.36^{+0.36}_{-0.33}$  & $6.6^{+8.0}_{-6.6}$  & $25 ^{+27} _{-9}$  &       0.78$\times10^{-10}$   &    0.927/152      \\
9                                  &  $7.7^{+4.1} _{-3.7}$   & $1.67^{+0.25}_{-0.25}$  & $6.8^{+5.6}_{-6.8}$  & $45 ^{+71} _{-27}$ &       1.57$\times10^{-10}$   &    0.872/223     \\
10                                 &  $5.9^{+4.2} _{-4.1}$   & $1.43^{+0.45}_{-0.30}$  & $6.1^{+7.5}_{-6.1}$  & $22 ^{+27} _{-7}$  &       1.27$\times10^{-10}$   &    0.793/224     \\
11                                 &  $5.7^{+3.1} _{-3.3} $  & $1.64^{+0.19}_{-0.19}$  & $6.1^{+1.7}_{-6.1}$  & $35 ^{+25} _{-12}$ &       1.74$\times10^{-10}$   &    0.991/268      \\
12                                 &  $4.7\pm{4.0} $         & $1.49^{+0.26}_{-0.23}$  & $5.8^{+2.5}_{-5.8}$  & $28 ^{+23} _{-9}$  &       1.11$\times10^{-10}$   &    0.868/215      \\
13                                 &  $3.0^{+4.1} _{-3.0}$   & $1.27^{+0.30}_{-0.22}$  & $7.0^{+2.4}_{-1.0}$  & $19 ^{+9} _{-4}$   &       1.38$\times10^{-10}$   &    0.963/325      \\  
14                                 &  $5.6^{+2.8} _{-2.5}$   & $1.49^{+0.20}_{-0.17}$  & $6.8^{+1.9}_{-1.2}$  & $26 ^{+12} _{-6}$  &       2.36$\times10^{-10}$   &    0.958/471      \\
15                                 &  $5.0^{+2.5} _{-2.0}$   & $1.38^{+0.23}_{-0.14}$  & $6.3^{+2.6}_{-0.8}$  & $23 ^{+8} _{-4}$  &       2.93$\times10^{-10}$   &    0.971/546      \\
\hline
\end{tabular}
\footnotesize{\\
$^a$ Power-law photon index. \\
$^b$ Flux corrected for the absorption. \\ 
}
\end{center}
\end{table*}

        \subsection{Intensity-selected spectroscopy}
        \label{sec:intsel}

Although there is no strong evidence of a trend in the hardness ratio (Fig.~\ref{fig:nulchr_int}, H/S=12--78 keV / 3-12 keV) with the source flux, nevertheless we performed an intensity-selected spectroscopy to investigate the  eventual presence of absorption features compatible with \cyclabs.
We extracted intensity selected spectra, after binning the net source light curve on the pulsar spin period, to avoid source variability due to the X-ray pulsations. 
Four X-ray intensity ranges were considered, to get comparable statistics, with net source count rate in the following intervals (shown in Fig.~\ref{fig:nulc}):  below 1~count~s$^{-1}$,  1--3~count~s$^{-1}$, 3--5~count~s$^{-1}$ and
above 5~count~s$^{-1}$, resulting into net exposure times of 13.5~ks, 17~ks, 9.7~ks and 3.5~ks, respectively in the four spectra. 
In Table~\ref{tab:intsel} we have listed the spectral parameters obtained with PLCUT model, confirming no variability with the X-ray flux, within the uncertainties.
No significant absorption lines were present.

\begin{table*}
\caption{Intensity-selected spectroscopy (see Sect.~\ref{sec:intsel}).}
\label{tab:intsel}
\vspace{0.0 cm}
\begin{center}
\begin{tabular}{lcccccccc} \hline
 \hline
\noalign {\smallskip}
Count rate range    &  N$_{\rm H}$            &   $\Gamma^a$            &   E$_{cut}$           &   E$_{fold}$      &    UF$_{1-100~keV}^b$      &  L$_{1-100~keV}$                & $\chi^{2}_{\nu}$/dof  & Net exposure\\   
  (count s$^{-1}$)  & ($10^{22}$~cm$^{-2}$)   &                         &   (keV)               &     (keV)         &   (erg~cm$^{-2}$~s$^{-1}$)  & (10$^{36}$~erg~s$^{-1}$)   &                    & (ks) \\     
\hline
\noalign {\smallskip}
$<$1                &  $5.3^{+2.2} _{-2.1}$   & $1.44^{+0.15}_{-0.16}$  & $6.3^{+1.1}_{-1.1}$   & $23  ^{+7} _{-5}$  &     0.70$\times10^{-10}$   & $0.35 ^{+0.17} _{-0.11}$ &  1.048/538    & 13.5 \\
1-3                 &  $4.6^{+1.4} _{-1.4}$   & $1.45^{+0.11}_{-0.10}$  & $6.5^{+1.0}_{-0.5}$   & $25 ^{+5} _{-4}$   &     1.31$\times10^{-10}$   & $0.66 ^{+0.32} _{-0.20}$ & 0.920/746    &  17.0 \\
3-5                 &  $6.6^{+1.4} _{-1.2}$   & $1.50^{+0.11}_{-0.08}$  & $6.5^{+1.0}_{-0.4}$   & $27 ^{+5} _{-3}$   &     2.90$\times10^{-10}$   & $1.47 ^{+0.70} _{-0.46}$ &  1.027/800    &  9.7 \\  
$>5$                &  $5.0^{+1.6} _{-1.6}$   & $1.33^{+0.12}_{-0.10}$  & $6.5 ^{+0.8} _{-0.5}$ & $21 ^{+4} _{-3}$   &     4.75$\times10^{-10}$   & $2.40 ^{+1.15} _{-0.74}$ &  1.019/681    &  3.5 \\
\hline
\end{tabular}
\footnotesize{\\
$^a$ Power-law photon index. \\
$^b$ Flux corrected for the absorption. \\ 
}
\end{center}
\end{table*}

\subsection{Spin-phase-resolved spectroscopy}
\label{sec:phasesel}
 
\begin{figure}
\includegraphics[scale=0.43,angle=0]{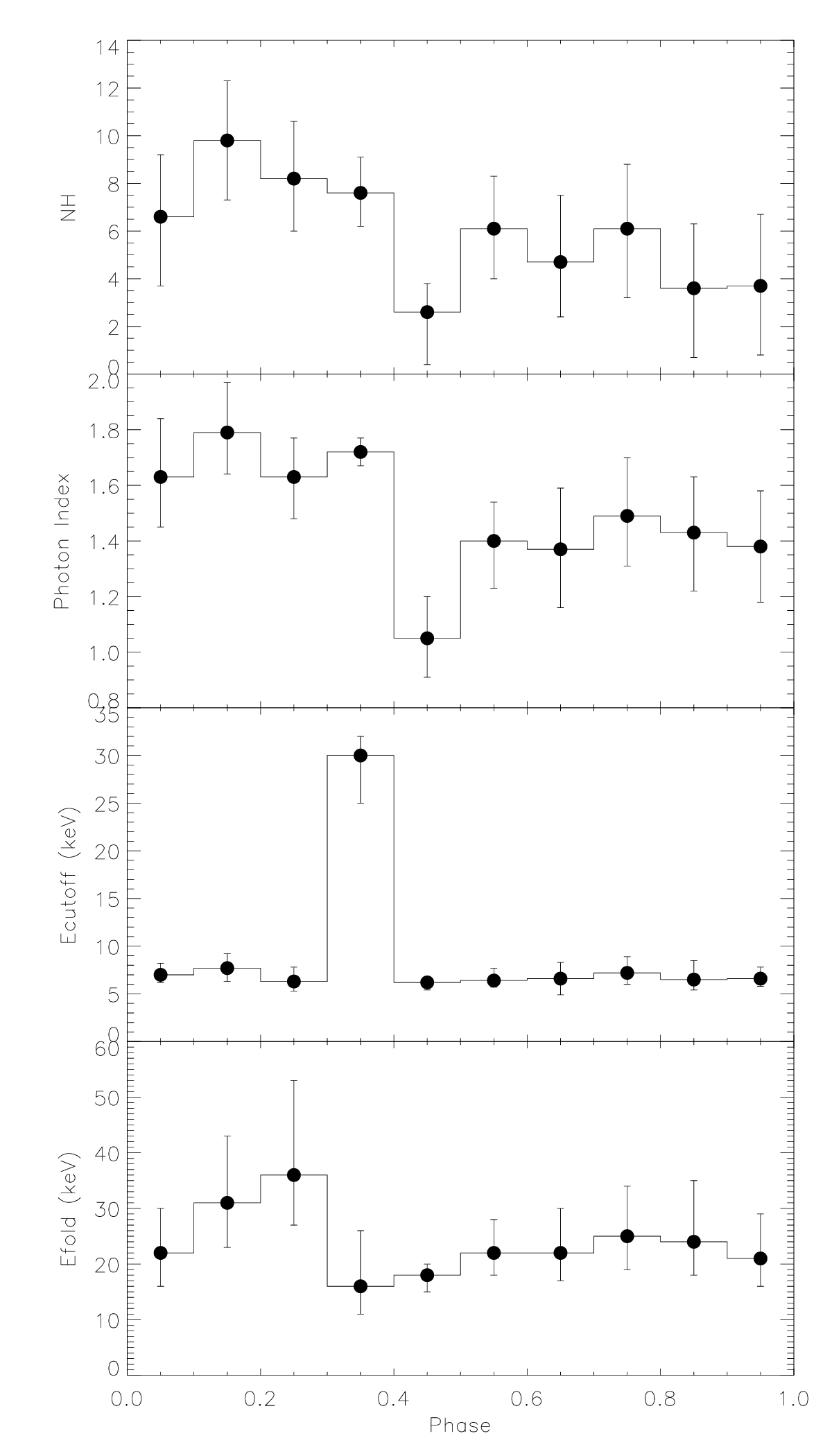} 
\caption{Spin-phase resolved spectroscopy over the pulse, obtained folding the whole \nustar\ observation on the spin period (the parameters are reported in Table~\ref{tab:spin_spec}). In the top panel, the absorption column density is in units of 10$^{22}$~cm$^{-2}$.
}
\label{fig:spin_spec}
\end{figure}

The source shows evidence of variability of the spectral hardness along the pulse profile  (Fig.~\ref{fig:pulseprofile_entire}). 
This behavior suggested to perform a spin-phase resolved spectroscopy.
The results are reported in Table~\ref{tab:spin_spec} and shown in Fig.~\ref{fig:spin_spec}.
Harder X-ray emission is found at spin phases $\Delta\phi$=0.3-0.5, preceding the main peak of the 3-12 keV profile, 
while the softest emission appears in the range $\Delta\phi$=0.0-0.2 (Fig.~\ref{fig:pulseprofile_entire}),
The spectral analysis indicates that this hardening can be explained by two different reasons (Fig.~\ref{fig:spin_spec}): 
in the spin phase interval $\Delta\phi$=0.3-0.4 the power-law extends up to very high energies 
(high E$_{cut}$ compared with other spin phases), while in  the interval $\Delta\phi$=0.4-0.5, the flattest power-law is obtained.

%
\begin{table*}
\caption{Spin-phase-resolved spectroscopy of the overall \nustar\ spectrum (see Sect.~\ref{sec:phasesel}). }\label{tab:spin_spec}
\vspace{0.0 cm}
\begin{center}
\begin{tabular}{lcccccc} \hline
 \hline
\noalign {\smallskip}
$\Delta\phi$                            &   N$_{\rm H}$            &   $\Gamma^a$            &   E$_{cut}$           &   E$_{fold}$      &    F$_{3-78~keV}^b$          & $\chi^{2}_{\nu}$/dof  \\   
                                        & ($10^{22}$~cm$^{-2}$)    &                         &   (keV)               & (keV)             &   (erg~cm$^{-2}$~s$^{-1}$)   &                      \\     
\hline
\noalign {\smallskip}
0.0-0.1                                  &  $6.6^{+2.6} _{-2.9}$   & $1.63^{+0.18}_{-0.21}$  & $7.0^{+1.2}_{-0.8}$  & $22  ^{+8} _{-6}$  &       1.14$\times10^{-10}$   &    0.994/408      \\
0.1-0.2                                  &  $9.8\pm{2.5} $         & $1.79^{+0.15}_{-0.18}$  & $7.7^{+1.5}_{-1.4}$  & $31 ^{+12} _{-8}$  &       1.43$\times10^{-10}$   &    1.117/458      \\
0.2-0.3                                  &  $8.2^{+2.4} _{-2.2}$   & $1.63^{+0.15}_{-0.14}$  & $6.3^{+1.5}_{-1.0}$  & $36 ^{+17} _{-9}$  &       1.67$\times10^{-10}$   &    1.107/483      \\                                
0.3-0.4                                  &  $7.6^{+1.5} _{-1.4}$   & $1.72^{+0.05}_{-0.05}$  & $30 ^{+2}  _{-5}$    & $16 ^{+10} _{-5}$  &       1.71$\times10^{-10}$   &    0.896/506      \\
0.4-0.5                                  &  $2.6^{+1.2} _{-2.2}$   & $1.05^{+0.14}_{-0.15}$  & $6.2^{+0.6}_{-0.8}$  & $18 ^{+2} _{-3}$   &       1.84$\times10^{-10}$   &    1.033/550      \\
0.5-0.6                                  &  $6.1^{+2.2} _{-2.1}$   & $1.40^{+0.17}_{-0.14}$  & $6.4^{+1.3}_{-0.7}$  & $22 ^{+6} _{-4}$   &       1.91$\times10^{-10}$   &    0.926/550      \\
0.6-0.7                                  &  $4.7^{+2.8} _{-2.3}$   & $1.37^{+0.21}_{-0.22}$  & $6.6^{+1.7}_{-1.7}$  & $22 ^{+8} _{-5}$   &       1.68$\times10^{-10}$   &    0.942/507      \\
0.7-0.8                                  &  $6.1^{+2.7} _{-2.9}$   & $1.49^{+0.18}_{-0.21}$  & $7.2^{+1.7}_{-1.2}$  & $25 ^{+9} _{-6}$   &       1.35$\times10^{-10}$   &    0.858/443      \\
0.8-0.9                                  &  $3.6^{+2.7} _{-2.9}$   & $1.43^{+0.21}_{-0.20}$  & $6.5^{+2.0}_{-1.1}$  & $24 ^{+11} _{-6}$  &       1.15$\times10^{-10}$   &    0.875/372     \\
0.9-1.0                                  &  $3.7^{+3.0} _{-2.9}$   & $1.38^{+0.20}_{-0.20}$  & $6.6^{+1.2}_{-0.8}$  & $21 ^{+8}  _{-5}$  &       1.13$\times10^{-10}$   &    1.047/374     \\
\hline
\end{tabular}
\footnotesize{\\
$^a$ Power-law photon index. \\
$^b$ Flux not corrected for the absorption. \\ 
}
\end{center}
\end{table*}
%

\subsection{Spin-phase-resolved spectroscopy of intensity selected spectra}
\label{sec:intsel_spinphase}

\begin{figure*}
\includegraphics[scale=0.37,angle=0]{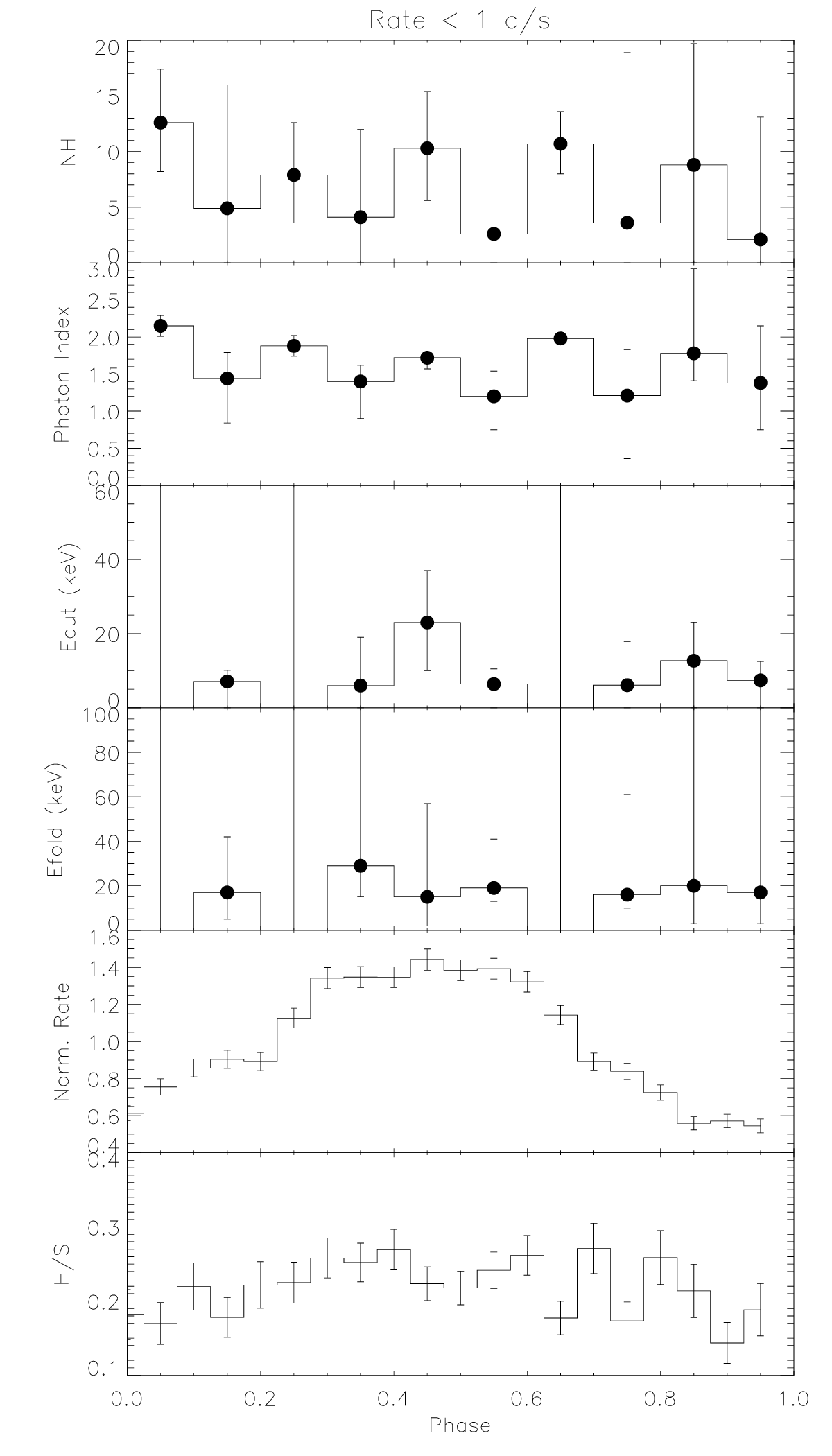} 
\includegraphics[scale=0.37,angle=0]{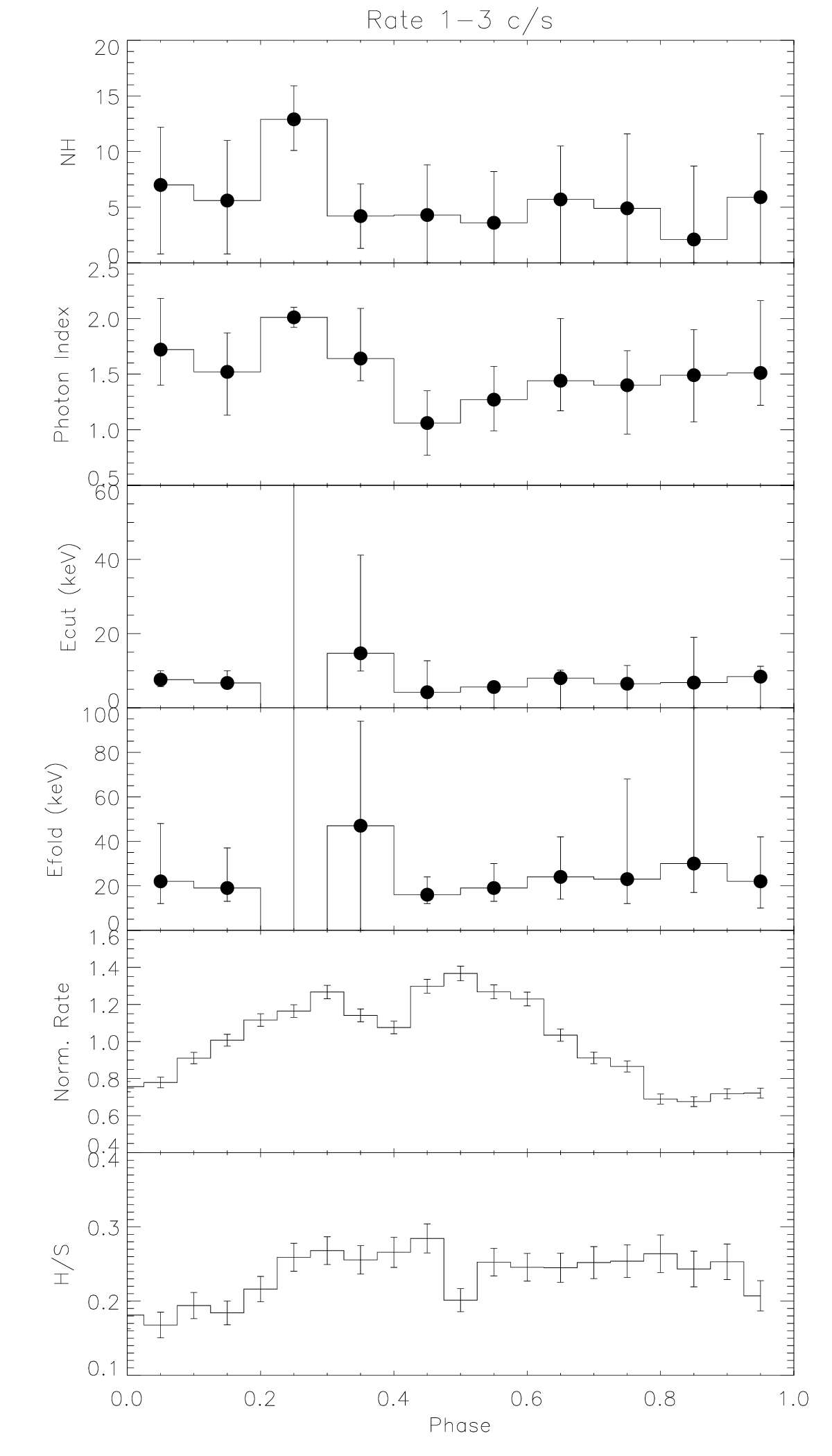} \\
\vspace{-0.3cm}
\includegraphics[scale=0.37,angle=0]{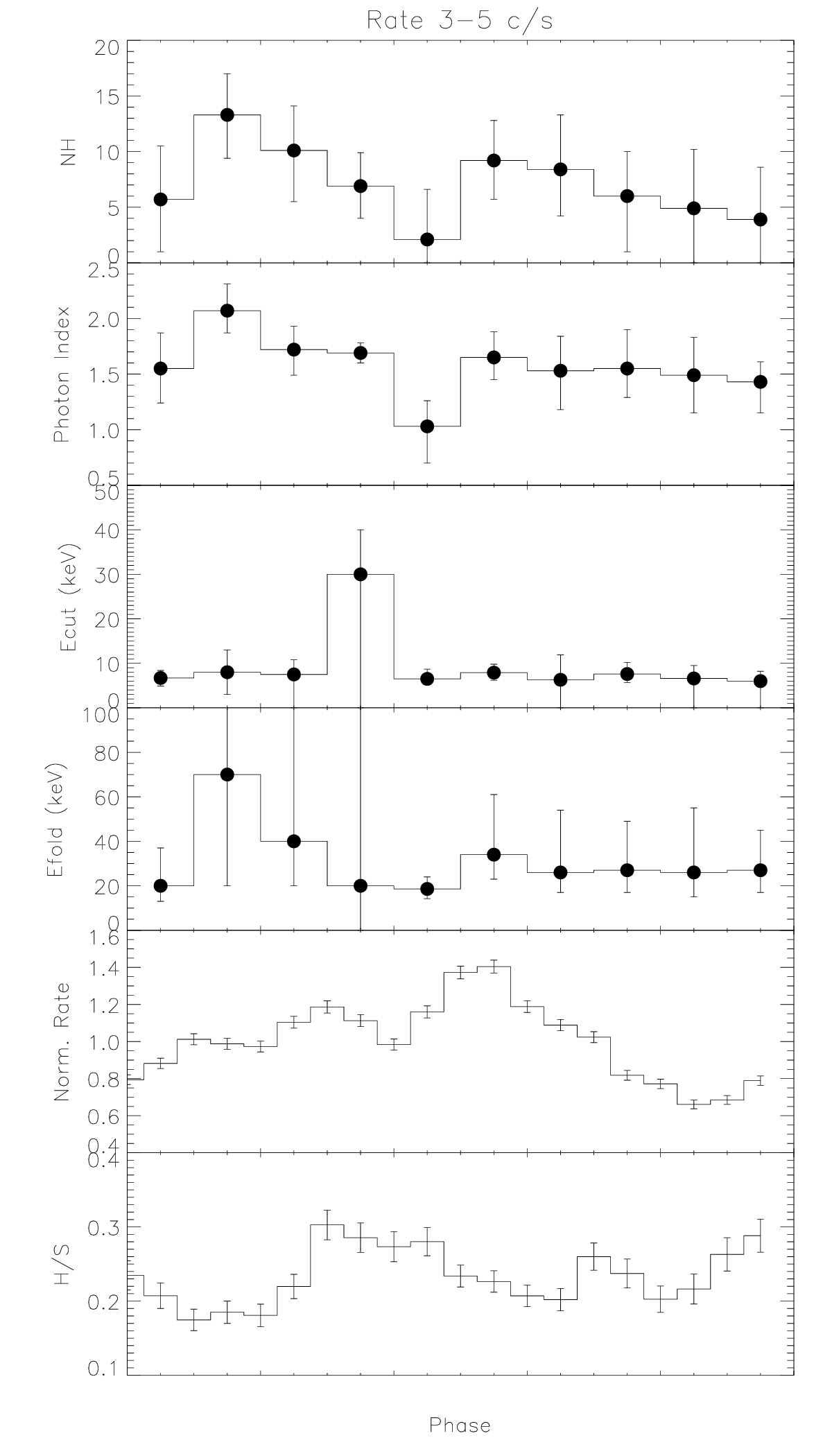} 
\includegraphics[scale=0.37,angle=0]{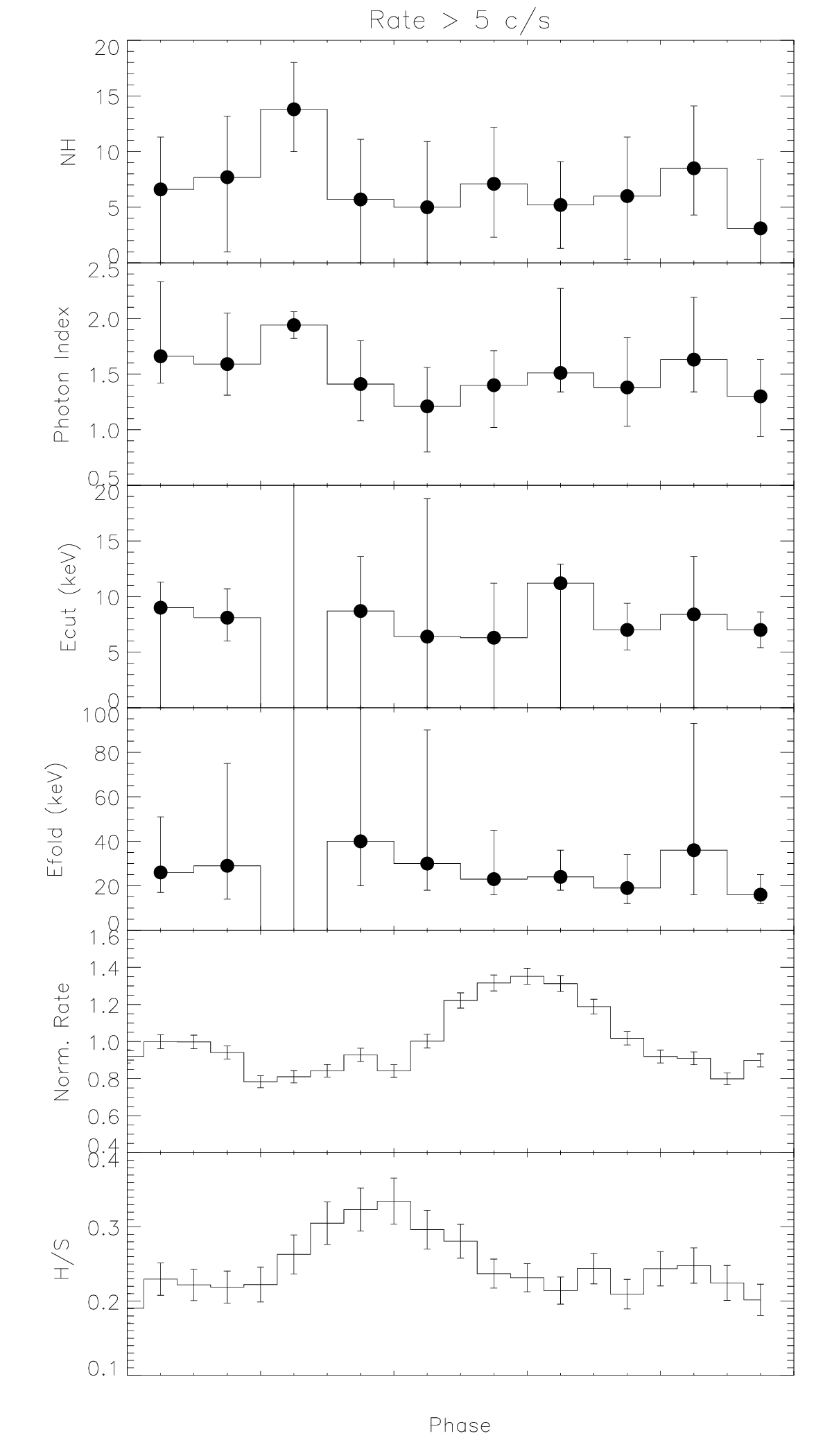} \\
\vspace{-0.3cm}
\caption{Spin-phase selected spectroscopy
in four intensity states, as reported in Table~\ref{tab:intsel_spinphase}: from top to bottom, from left to right, the intensity states are below 1~count~s$^{-1}$,  1--3~count~s$^{-1}$, 3--5~count~s$^{-1}$ and above 5~count~s$^{-1}$.
In the bottom panels, the intensity-selected pulse profiles (3-78 keV) are reported, together with the hardness ratio (H/S) of hard pulse profile (H=12-78 keV) to the soft pulse profile (S=3-12 keV).
}
\label{fig:intsel_spinphase}
\end{figure*}

In Table~\ref{tab:intsel_spinphase} and Fig.~\ref{fig:intsel_spinphase} we report on the spin-phase selected spectroscopy performed within the four intensity states, adopting the same model and ten spin phase intervals per period. 
Given the shorter exposure time compared with previous spectroscopy, the energy range covered by these spectra is limited to the band 3-50 keV, with the only exception of the lowest luminosity state at the minimum of the pulse profile, where the spectrum extends up to 30 keV.  
No evidence of significant absorption features was obtained.
In a few cases, the parameters of the high energy cutoff component (E$_{cut}$ and E$_{fold}$) were unconstrained (and are indicated by vertical lines in Fig.~\ref{fig:intsel_spinphase}), and a single absorbed power-law was an acceptable deconvolution of the X-ray continuum emission.
%

\begin{table*}
\caption{Results of the spin-phase-resolved spectroscopy of intensity selected spectra (see Sect.~\ref{sec:intsel_spinphase}).}
\label{tab:intsel_spinphase}  
\vspace{0.0 cm}
\begin{center}
\begin{tabular}{lcccccc} \hline   
 \hline
\noalign {\smallskip}
Rate $<$1~count~s$^{-1}$ &                        &                         &                       &                   &                               &                     \\
\cline{1-1}
$\Delta\phi$           &   N$_{\rm H}$            &   $\Gamma^a$            &   E$_{cut}$           &   E$_{fold}$      &    UF$_{1-100~keV}^b$         & $\chi^{2}_{\nu}$/dof  \\   
                       & ($10^{22}$~cm$^{-2}$)    &                         &   (keV)               &   (keV)           &   (erg~cm$^{-2}$~s$^{-1}$)    &                      \\     
\hline
\noalign {\smallskip}
0.0-0.1                &  $12.6^{+4.8} _{-4.4}$  & $2.15^{+0.14}_{-0.14}$  & $-$                  &  $-$               &       0.90$\times10^{-10}$   &    1.017/43      \\
0.1-0.2                &  $4.9^{+11.1} _{-4.9}$  & $1.44^{+0.60}_{-0.35}$  & $7.1^{+3.0}_{-7.1}$  & $17 ^{+25} _{-12}$ &       0.71$\times10^{-10}$   &    0.615/52      \\
0.2-0.3                &  $7.9^{+4.7} _{-4.3}$   & $1.88^{+0.14}_{-0.14}$  &  $-$                 &  $-$               &       1.37$\times10^{-10}$   &    1.017/74      \\ 
0.3-0.4                &  $4.1^{+7.9} _{-4.1}$   & $1.40^{+0.50}_{-0.22}$  & $6  ^{+13} _{-6}$    & $29 ^{+160} _{-14}$&       1.16$\times10^{-10}$   &    0.977/97      \\
0.4-0.5                &  $10.3^{+5.1} _{-4.7}$  & $1.72^{+0.15}_{-0.07}$  & $23^{+14}_{-13}$     & $15 ^{+42} _{-13}$ &       1.21$\times10^{-10}$   &    0.918/102      \\
0.5-0.6                &  $2.6^{+6.9} _{-2.6}$   & $1.20^{+0.45}_{-0.34}$  & $6.4^{+4.1}_{-6.4}$  & $19 ^{+22} _{-6}$  &       1.09$\times10^{-10}$   &    0.938/107      \\
0.6-0.7                &  $10.7^{+2.9} _{-2.7}$  & $1.98^{+0.08}_{-0.08}$  &  $-$                 &  $-$               &       1.30$\times10^{-10}$   &    1.167/82      \\
0.7-0.8                &  $3.6^{+15.3} _{-3.6}$  & $1.21^{+0.85}_{-0.62}$  & $6.1^{+11.7}_{-6.1}$ & $16 ^{+45} _{-6}$  &       0.58$\times10^{-10}$   &    1.044/56      \\
0.8-0.9                &  $8.8^{+10.9} _{-8.8}$  & $1.78^{+0.37}_{-1.14}$  &$12.7^{+10.4}_{-12.7}$& $20^{+350} _{-17}$ &       0.58$\times10^{-10}$   &    0.998/41     \\
0.9-1.0                &  $2.1^{+11.0} _{-2.1}$  & $1.38^{+0.63}_{-0.77}$  & $7.4^{+5.1}_{-7.4}$  & $17 ^{+108} _{-14}$&       0.46$\times10^{-10}$   &    0.831/34     \\
\hline
\noalign {\smallskip}
Rate 1--3~count~s$^{-1}$ &                        &                         &                       &                   &                               &                     \\
\cline{1-1}
$\Delta\phi$           &   N$_{\rm H}$            &   $\Gamma^a$            &   E$_{cut}$           &   E$_{fold}$      &    UF$_{1-100~keV}^b$         & $\chi^{2}_{\nu}$/dof  \\   
                       & ($10^{22}$~cm$^{-2}$)    &                         &   (keV)               &   (keV)           &   (erg~cm$^{-2}$~s$^{-1}$)    &                      \\     
\hline
\noalign {\smallskip}
0.0-0.1                &  $7.0^{+5.2} _{-6.2}$  & $1.72^{+0.32}_{-0.46}$   & $7.6^{+2.4}_{-1.9}$  & $22^{+26}_{-10}$   &       1.07$\times10^{-10}$   &    1.029/130      \\
0.1-0.2                &  $5.6^{+5.4} _{-4.8}$  & $1.52^{+0.39}_{-0.35}$   & $6.7^{+3.3}_{-1.6}$  & $19 ^{+18} _{-6}$  &       1.40$\times10^{-10}$   &    1.090/173      \\
0.2-0.3                & $12.9^{+3.0} _{-2.8}$  & $2.01^{+0.09}_{-0.09}$   &  $-$                 &  $-$               &       2.56$\times10^{-10}$   &    1.106/193      \\ 
0.3-0.4                &  $4.2^{+2.9} _{-2.9}$  & $1.64^{+0.20}_{-0.45}$  &$14.7 ^{+26.5} _{-4.8}$& $47 ^{+47} _{-47}$ &       1.96$\times10^{-10}$   &    0.849/194      \\
0.4-0.5                & $4.3^{+4.5} _{-4.3}$   & $1.06^{+0.29}_{-0.29}$   & $4.2^{+8.5}_{-4.2}$  & $16 ^{+8} _{-4}$   &       1.69$\times10^{-10}$   &    0.904/205      \\
0.5-0.6                &  $3.6^{+4.6} _{-3.6}$   & $1.27^{+0.28}_{-0.30}$  & $5.6^{+1.6}_{-5.6}$  & $19 ^{+11} _{-6}$  &       1.66$\times10^{-10}$   &    0.905/189      \\
0.6-0.7                &  $5.7^{+4.8} _{-5.7}$  & $1.44^{+0.27}_{-0.56}$   & $8.0^{+2.2}_{-8.0}$  & $24^{+18}_{-10}$   &       1.52$\times10^{-10}$   &    0.919/168     \\
0.7-0.8                &  $4.9^{+6.7} _{-4.9}$  & $1.40^{+0.44}_{-0.31}$   & $6.5^{+4.9}_{-6.5}$  & $23 ^{+45} _{-11}$ &       1.24$\times10^{-10}$   &    0.720/135      \\
0.8-0.9                &  $2.1^{+6.6} _{-2.1}$  & $1.49^{+0.42}_{-0.41}$   & $6.8^{+12.2}_{-6.8}$ & $30^{+72} _{-13}$  &       0.99$\times10^{-10}$   &    1.008/110     \\
0.9-1.0                &  $5.9^{+5.7} _{-5.9}$  & $1.51^{+0.29}_{-0.65}$   & $8.4^{+2.8}_{-8.4}$  & $22 ^{+20} _{-12}$ &       1.01$\times10^{-10}$   &    1.009/117     \\
\hline
\noalign {\smallskip}
Rate 3--5~count~s$^{-1}$ &                        &                         &                       &                   &                               &                     \\
\cline{1-1}
$\Delta\phi$           &   N$_{\rm H}$            &   $\Gamma^a$            &   E$_{cut}$           &   E$_{fold}$      &    UF$_{1-100~keV}^b$         & $\chi^{2}_{\nu}$/dof  \\   
                       & ($10^{22}$~cm$^{-2}$)    &                         &   (keV)               &   (keV)           &   (erg~cm$^{-2}$~s$^{-1}$)    &                      \\     
\hline
\noalign {\smallskip}
0.0-0.1                &  $5.7^{+4.8} _{-4.7}$  & $1.55^{+0.31}_{-0.32}$   & $6.7^{+1.7}_{-1.8}$  & $20^{+17}_{-7}$    &       2.34$\times10^{-10}$   &    0.872/166      \\
0.1-0.2                & $13.3^{+3.7} _{-3.9}$  & $2.07^{+0.20}_{-0.24}$   & $8.0^{+5.0}_{-5.0}$  &$70 ^{+100} _{-50}$ &       3.84$\times10^{-10}$   &    0.998/192      \\
0.2-0.3                & $10.1^{+4.0} _{-4.6}$  & $1.72^{+0.23}_{-0.21}$   & $7.5^{+3.3}_{-7.5}$  & $40 ^{+70} _{-20}$ &       3.70$\times10^{-10}$   &    1.032/201      \\ 
0.3-0.4                &  $6.9^{+3.0} _{-2.9}$  & $1.69^{+0.09}_{-0.09}$   & $30 ^{+10} _{-30}$   &$20 ^{+600} _{-20}$ &       3.41$\times10^{-10}$   &    0.959/201      \\
0.4-0.5                & $2.1^{+4.5} _{-2.1}$   & $1.03^{+0.33}_{-0.23}$   & $6.5^{+2.2}_{-1.1}$  & $18.6^{+5.4}_{-4.4}$ &     3.23$\times10^{-10}$   &    0.951/220      \\
0.5-0.6                & $9.2^{+3.6} _{-3.5}$   & $1.65^{+0.20}_{-0.23}$   & $7.9^{+1.9}_{-1.7}$  & $34 ^{+27} _{-11}$ &       4.00$\times10^{-10}$   &    1.054/243      \\
0.6-0.7                &  $8.4^{+4.9} _{-4.2}$  & $1.53^{+0.35}_{-0.31}$   & $6.3^{+5.6}_{-6.3}$  & $26^{+28}_{-9}$    &       3.19$\times10^{-10}$   &    0.717/205     \\
0.7-0.8                &  $6.0^{+4.0} _{-5.0}$  & $1.55^{+0.26}_{-0.35}$   & $7.6^{+2.6}_{-1.9}$  & $27 ^{+22} _{-10}$ &       2.58$\times10^{-10}$   &    0.794/169      \\
0.8-0.9                &  $4.9^{+5.3} _{-4.9}$  & $1.49^{+0.34}_{-0.34}$   & $6.6^{+2.9}_{-6.6}$  & $26^{+29} _{-11}$  &       2.09$\times10^{-10}$   &    0.912/142     \\
0.9-1.0                &  $3.9^{+4.7} _{-3.9}$  & $1.43^{+0.28}_{-0.18}$   & $6.0^{+2.2}_{-6.0}$  & $27 ^{+18} _{-10}$ &       2.25$\times10^{-10}$   &    0.960/153     \\
\hline
\noalign {\smallskip}
Rate $>$5~count~s$^{-1}$ &                        &                         &                       &                   &                               &                     \\
\cline{1-1}
$\Delta\phi$           &   N$_{\rm H}$            &   $\Gamma^a$            &   E$_{cut}$           &   E$_{fold}$      &    UF$_{1-100~keV}^b$         & $\chi^{2}_{\nu}$/dof  \\   
                       & ($10^{22}$~cm$^{-2}$)    &                         &   (keV)               &   (keV)           &   (erg~cm$^{-2}$~s$^{-1}$)    &                      \\     
\hline
\noalign {\smallskip}
0.0-0.1                &  $6.6^{+4.7} _{-6.6}$  & $1.66^{+0.24}_{-0.67}$   & $9.0^{+2.3}_{-9.0}$  & $26^{+25}_{-9}$    &       4.58$\times10^{-10}$   &    0.993/130      \\
0.1-0.2                &  $7.7^{+5.5} _{-6.7}$  & $1.59^{+0.28}_{-0.46}$   & $8.1^{+2.6}_{-2.1}$  &$29 ^{+46} _{-15}$  &       4.62$\times10^{-10}$   &    1.098/131      \\
0.2-0.3                & $13.8^{+4.2} _{-3.8}$  & $1.94^{+0.12}_{-0.12}$   & $-$                  & $-$                &       6.04$\times10^{-10}$   &    0.906/114      \\
0.3-0.4                &  $5.7^{+5.4} _{-5.7}$  & $1.41^{+0.33}_{-0.39}$   & $8.7 ^{+4.9} _{-8.7}$&$40 ^{+70} _{-20}$  &       5.37$\times10^{-10}$   &    1.098/120      \\
0.4-0.5                & $5.0^{+5.9} _{-5.0}$   & $1.21^{+0.41}_{-0.35}$   & $6.4^{+12.4}_{-6.4}$ & $30^{+60}_{-12}$   &       5.81$\times10^{-10}$   &    0.991/143      \\
0.5-0.6                & $7.1^{+5.1} _{-4.8}$   & $1.40^{+0.38}_{-0.31}$   & $6.3^{+4.9}_{-6.3}$  & $23 ^{+22} _{-7}$  &       6.56$\times10^{-10}$   &    0.933/178      \\
0.6-0.7                &  $5.2^{+3.9} _{-3.9}$  & $1.51^{+0.17}_{-0.76}$   & $11.2^{+1.7}_{-11.2}$& $24^{+12}_{-6}$    &       6.33$\times10^{-10}$   &    1.084/174     \\
0.7-0.8                &  $6.0^{+5.3} _{-5.7}$  & $1.38^{+0.35}_{-0.45}$   & $7.0^{+2.4}_{-1.8}$  & $19 ^{+15} _{-7}$  &       4.61$\times10^{-10}$   &    1.009/148      \\
0.8-0.9                &  $8.5^{+5.6} _{-4.2}$  & $1.63^{+0.29}_{-0.56}$   & $8.4^{+5.2}_{-8.4}$  & $36^{+57} _{-20}$  &       4.53$\times10^{-10}$   &    0.962/125     \\
0.9-1.0                &  $3.1^{+6.2} _{-3.1}$  & $1.30^{+0.36}_{-0.33}$   & $7.0^{+1.6}_{-1.6}$  & $16 ^{+9} _{-4}$   &       3.66$\times10^{-10}$   &    0.958/119     \\
\hline
\end{tabular}
\footnotesize{\\
$^a$ Power-law photon index. \\
$^b$ Flux corrected for the absorption. \\ 
}
\end{center}
\end{table*}

\section{Discussion}

We have reported here on the X-ray properties shown by the SFXT pulsar \src\  during the periodic outburst occurred in June, 2017, 
observed with the \nustar\ satellite (3-78 keV) at its maximum brightness.
Monitoring observations with \sw/XRT (2-8 keV) have confirmed the rise to the outburst peak and the shortness of the flaring activity phase,  
in agreement with previous ones, where the X-ray luminosity was above 10$^{34}$ erg s$^{-1}$ for less than 
10 days \citep{Romano2007, Romano2009}. 
   
The \nustar\ observation resulted into a net exposure time of about 44 ks, but since \nustar\ operates in a low orbit, with several observational gaps, the total temporal coverage of the X-ray activity was about one day,
a time span corresponding to the typical duration of the peak of the outburst  in \src.

The \nustar\ light curve showed a flaring activity with an ample range of variability, reaching a factor of 100.
Although the light curve is segmented in 15 satellite revolutions (so that the  temporal evolution of the X-ray variability cannot be entirely followed), it is possible to distinguish a few short ($\sim$hundred seconds) X-ray flares within single satellite orbits.
This huge flaring variability in outburst is the typical X-ray activity characterizing the  SFXTs as a class.

The X-ray spectrum was well modeled with an absorbed power-law, modified by a high energy cutoff, a phenomenological deconvolution frequently used to describe the spectra of accreting pulsars (e.g., \citealt{Coburn2002}), together with a faint emission line at 6.4 keV, consistent with neutral iron, usually observed in HMXBs and due to fluorescent emission in the companion wind \citep{Torrejon2010, Gimenez2015}.

This is the second time \nustar\ observes \src\ in outburst, but now with a doubled exposure time (i.e., $\sim$44 ks) with respect to the
2016 outburst reported by \citet{Sidoli2017}. 
In 2016 the outburst was followed also by \xmm, simultaneously with \nustar, implying a very sensitive coverage of the soft region of the spectrum (below 3 keV), that showed a soft excess which could be modeled equally 
well either with a hot blackbody or with an additional absorption model, covering only a fraction of the power-law emission at soft energies.

In the here reported 2017 outburst, the \sw/XRT observations were too short  to detect the source below 2 keV, and the 2-78 keV spectrum did not show evidence of a more complex model than a single absorbed power-law with a high energy roll-over. 

The \nustar\ spectra collected in 2016 and 2017, extracted from the whole exposures, were very similar. 
The re-analysis of the whole 2016 exposure (20 ks), 
showed a slightly less absorbed ((2.4$\pm{0.7})\times10^{22}$~cm$^{-2}$) harder power-law model
($\Gamma$=1.26$^{+0.10} _{-0.07}$), with a bit higher flux of F=1.6$\times10^{-10}$~erg~cm$^{-2}$~s$^{-1}$ 
(3-78, not corrected for the absorption) than in 2017.
The cutoff and e-folding energies, together with the parameters of the faint 
and narrow emission line from neutral iron, were compatible in the two outbursts.

During the spectroscopy of a flare of the 2016 outburst, we found \citep{Sidoli2017}
a hint (at 2.63~$\sigma$) of an absorption feature at 17 keV, probably a variable \cyclab\ which
needed confirmation. 
Unfortunately we did not find any evidence of  \cyclabs\ in any of the
spectra extracted from the 2017 observation. 
Indeed, we have searched  for significant absorption features by adopting many different spectral extractions: time-averaged spectrum, 
temporal-selected, spin-phase-selected and intensity-selected spectra (together with a spin-phase resolved spectroscopy at different 
X-ray intensity states). All these spectra were well fitted with the same model, with the exception of the faint K$_{\alpha}$ iron emission line, which was clearly detected only in the longest exposure, time-averaged spectrum. 
In none of them we found significant absorption features compatible with the presence of a \cyclab.

X-ray pulsations have been clearly detected at a periodicity \ps=187.0$\pm{0.12}$ s, consistent with past  observations \citep{Swank2007, Sidoli2007, Sidoli2017}, showing a remarkable stability of the pulsar rotation since its discovery, maybe suggestive of the fact that the equilibrium period has been reached.

The pulsed fraction showed an evolution with both the energy and the X-ray luminosity, increasing at harder energies, and decreasing at higher luminosity. 
The changing pulse shape is typically observed in accreting pulsars and has been investigated by many authors, especially in transient Be X-ray binaries, which span a large range of X-ray luminosity \citep{Parmar1989}.
An evolution in the pulse profile indicates a change in the radiation beam pattern, 
and is thought to be due to a changing accretion regime onto magnetized NS.
In particular, it has been investigated in the most luminous and transient X-ray pulsars crossing the critical luminosity threshold
at $\sim$10$^{37}$~erg~s$^{-1}$ \citep{Becker2012}. 
However, in \src\ the luminosity reached in outburst is always sub-critical, 
at $\sim$10$^{36}$~erg~s$^{-1}$ (1-100 keV).
Nevertheless, we observed a changing pulse pattern from a single, sinusoidal, peak at low luminosity (at an average 
L$_{\rm 1-100\,keV}$=3.5$\times10^{35}$~erg~s$^{-1}$, 
see Table~\ref{tab:intsel}) and in all energy ranges (3-12 keV, 12-78 keV and 3-78 keV), up to a double-peaked pattern at the highest luminosity range (at the peak of the brightest flares, 
L$_{\rm 1-100\,keV}$=2.4$\times10^{36}$~erg~s$^{-1}$), composed of a main and a secondary peak, with a main peak shifted with respect to the single-peak of the lowest luminosity state
by about $\Delta\phi$=+0.15.

A more complex pulse profile  is observed at intermediate luminosities (Fig.~\ref{fig:pulse_b}).  
In particular, at soft energies (3-12 keV), a notch  appears at spin phase $\phi$=0.4,
 in the profile extracted from the count rate range 1-3~count~s$^{-1}$ 
(average L$_{\rm 1-100\,keV}$=6.6$\times10^{35}$~erg~s$^{-1}$, Table~\ref{tab:intsel}),  
becoming deeper and broader towards higher luminosity states. 
At hard energies (12-78 keV) the pulse shape undergoes a less clear evolution with the luminosity, 
 but with a similar final result: the profile changes from a single-peaked (low luminosity) to a double-peaked profile at the highest luminosity state 
 (average L$_{\rm 1-100\,keV}$=2.4$\times10^{36}$~erg~s$^{-1}$), but here with a broader main peak than
the one observed in the energy range 3-12 keV. 

We note that a  pulse profile evolution was already observed   in \src\ using \xmm\ to investigate the 2006 outburst \citep{Sidoli2007}. 
In that occasion, the change between a single-peaked at low luminosity, 
to a double-peaked profile (at high luminosity) resulted in a larger phase shift of the main peak with respect to the single-peak at lower luminosity.
But it is worth noting that the low luminosity state probed by \xmm\ was fainter than the lowest luminosity state probed by \nustar: indeed,  rescaling luminosities in the same band (0.5-10 keV) and at the same distance of 6.5 kpc, we have found
that in 2006 the single-peaked pulse profile of the faint state was at a luminosity of 0.9$\times10^{34}$~erg~s$^{-1}$ (0.5-10 keV),
while with \nustar\ it was observed at an average luminosity of 2$\times10^{35}$~erg~s$^{-1}$ (0.5-10 keV).
Thus, we can conclude that the phase shift was larger in 2006 because the pulse profile was extracted at a lower luminosity with \xmm, than in 2017 using \nustar.

\subsection{X-ray pulse profile evolution in \src\ with luminosity: a physical explanation} 
\label{s:model}

The analysis of the pulse profile properties (time evolution with X-ray flux and time-resolved spectroscopy) suggests the following features that should be explained by a physical model:

\begin{enumerate}

\item The 3-12 keV pulse profile shape  changes with X-ray flux, from a single-peak sine-line form at low luminosities (a few $10^{35}$~erg~s$^{-1}$) to more complex double-peak profile at high luminosities (a few $10^{36}$~erg~s$^{-1}$), while the 12-78 keV pulse shape almost does not change (see Figs. 4-9).

\item The hardness ratio HR changes along with the pulse (Fig. 9) but keeps, on average, constant at different X-ray luminosities (Fig. 3).

\item The pulsed fraction $PF$ increases below $\sim 6\times 10^{35}$ erg~s$^{-1}$  (Fig. 6).

\item No CRSF has been found in the 3-78 keV spectral range (Fig. 7).

\end{enumerate}

At low and moderate accretion rates, the X-ray emission is generated by converting the accreting plasma kinetic energy into heat in a thin layer near the NS surface \citep{1969SvA....13..175Z,1973ApJ...184..271L}. 
The shape of the X-ray pulse profile and its evolution with luminosity are determined by the physical conditions above the NS polar cap (magnetic field strength, plasma density and temperature) and are insensitive to the mechanism triggering the source flaring activity.

\subsubsection{Photon scattering in accreting plasma
above one NS magnetic pole}

First, let us suppose that in \src\ we 
are observing only  X-ray pulsations from one NS magnetic pole, as suggested by the 12-78 keV pulse profile. 
This is possible for a given range of viewing angles of the NS spin and magnetic axes. In this case, the following physical explanation to the observed gross properties of the X-ray pulse evolution with luminosity can be proposed.

\cite{2013MNRAS.428..670S} discussed, in the context of the ‘off-states’ of the canonical wind-accreting X-ray pulsar Vela X-1, the pulse profile changes in low- and moderate-luminosity X-ray pulsars with luminosity. The pulse profile evolution is expected due to the possible transformation of a pencil X-ray beam produced by ordinary (O) photons at the NS polar caps into a more spherically-symmetric or even fan beam at high accretion rate due to photon scatterings in the accreting matter and possible vacuum polarization effects. This transition can occur above the X-ray luminosity 
$$
L_\dag \sim 3\times 10^{35}\mu_{30}^{-3/10} \mathrm{erg\,s}^{-1}
$$ 
Here $\mu=10^{30}\mu_{30}$~G~cm$^3$ is the NS dipole magnetic moment normalized to a canonical NS surface magnetic field of $10^{12}$~G. 

In this model, the change in the X-ray beam for photons above the cyclotron energy $E_{cyc}$ from pencil-like to fan-like as the X-ray luminosity increases leads to the X-ray pulse phase shift by about 90 degrees, as indeed is the case for Vela X-1 with $E_{cyc}\approx 20$~keV \citep{2011A&A...529A..52D}. 

In \src\, however, only a slight pulse phase shift has been detected (\citealt{Sidoli2007}, the \xmm\ observations, and the present paper). The unchanged mean hardness ratio HR (property 2 of Sect.~\ref{s:model}) suggests that the observed absorption feature in the X-ray pulse observed at high luminosities is due to scatterings of O-photons generated near the polar cap of the magnetized NS. The smallness in the X-ray pulse phase shift in the range 3-78 keV observed by \nustar\ can be interpreted by the lack of transformation of the pencil X-ray diagram in \src\  with luminosity. This is possible if there is no conversion of ordinary to extraordinary (X) photons in the strong magnetic field expected for photons below $E_{cyc}$ (see the discussion in \citealt{2013MNRAS.428..670S}), likely pointing to a value of $E_{cyc}$ below three keV, i.e. in the observed X-ray band 3-78 keV we observe only ordinary photons. As $E_{cyc}$ in \src\  has not been measured, it may seem difficult to make robust conclusions. 

However, we can make use of the fact that the X-ray pulse period of \src\  stays constant at $P_*=187$~s, suggesting its equilibrium value. In this case, to eliminate the unknown NS magnetic field, we can use the expression for the equilibrium period for quasi-spherical settling accretion (which appears to be the case for SFXTs, see \cite{Shakura2014} for more detail), 
$$
P_{eq}\approx 1000 F(e) \mu_{30}^{12/11} (P_b/10\,\mathrm{d}) \dot M_{16}^{-4/11} v_8^4\, \mathrm{s}.
$$ 
Here $P_b$ is the binary orbital period, $v=10^8v_8\,\mathrm{cm\,s}^{-1}$ is the optical stellar wind velocity near the NS orbital location, $\dot M=10^{16} \dot M_{16}\, \mathrm{g\, s}^{-1}$ is the mass accretion rate related to the X-ray accretion luminosity as $L_X=0.1 \dot M c^2$ ($c$ is the velocity of light). The factor $F(e)<1$ takes into account the possible reduction in the value of $P_{eq}$ due to the orbital eccentricity. For $e$ up to $\sim 0.8$ this factor can be $F(e)\sim 0.1-0.01$ \citep{2019MNRAS.485..851Y}, depending on the (unknown in \src\ ) wind radiation acceleration parameters and orbital eccentricity.

The scattering of O-photons generated in 
the NS polar cap at low accretion rates mostly occurs in the freely falling matter canalized by the NS magnetic field. For a NS dipole field, the optical depth in the accreting matter for ordinary photons above $E_{cyc}$ with the Thomson scattering cross-section is $\tau_v\simeq 3 (R_A/10^9 \mathrm{cm})^{1/2}\dot M_{16}$ \citep{1973ApJ...184..271L}. Inserting here the expression for the Alfven radius, $R_A \sim 1.4\times  10^9\mu_{30}^{6/11}\dot M_{16}^{-2/11}\,\mathrm{cm}$, for the quasi-spherical settling accretion \citep{Shakura2012} and eliminating the unknown NS magnetic field through the equilibrium spin period, we arrive at the optical depth
$$
\tau_v\simeq 3.6 (P_*/10^3 \mathrm{s})^{1/4}F(e)^{-1/4}(P_b/10\,\mathrm{d})^{-1/4} \dot M_{16}v_8^{-1}\,.
$$
For \src\ $P_*=187$~s and the binary orbital period is $P_b\approx 165$ days, so we get  $\tau_v\simeq 1.2 \dot M_{16}F(e)^{-1/4}v_8^{-1}$. We note that the uncertain reduction of $P_{eq}$ due to orbital eccentricity $F(e)<1$ only increases this estimate. 

As the wind velocity can hardly be higher than 1000 km~s$^{-1}$ (especially considering the likely origin of the outbursts in this source near the periastron passages, i.e., closer to the optical star where the wind velocity is lower), we can conclude that at high X-ray luminosities of \src\  observed (a few $10^{36}$ erg~s$^{-1}$ ), the optical depth in the accreting matter above the NS polar caps is indeed substantial for the O-photon scatterings. 

Therefore, the properties 1 and 2 (Sect.~\ref{s:model}) of the observed pulse profiles in \src\  listed above can be naturally explained by this model. We note that the pulse profile evolution found in the previous \xmm\ observations (see Fig.~3 in \citealt{Sidoli2007}) can also be explained by the scattering in the accreting material as described above. Indeed, in those observations, the mean hardness ratio HR also does not change at different X-ray luminosity states, and the pulse shape demonstrates strong scattering absorption feature at the bright state of the source. 

In the proposed model, we have eliminated the NS magnetic field by expressing it through the observed NS spin and binary orbital period. The non-observation of the CRSF feature in the \nustar\ spectrum (property 4 above) may imply either a rather strong NS magnetic field, $\mu_{30}\gtrsim 10$, or, oppositely, a weak field, $\mu_{30}<1$. The low NS magnetic field seems more likely because, for the high field, the vacuum polarization effects leading to the O-X photon transformation can be significant. The scattering cross-section of the X-photons is reduced below $E_{cyc}$: $\sigma_\perp\sim \sigma_T(E/E_{cyc})^2$, and so does the optical depth in the accreting matter above the NS polar caps. However, the scattering absorption feature in the pulse profiles signaling a large scattering optical depth at high luminosities has been observed in both \xmm\ \citep{Sidoli2007} and the \nustar\ pulse profiles (this paper).

Scattering of the O-photons in the optically thick accreting matter above the polar caps also explains why the pulse fraction increases with decreasing X-ray luminosity (see Fig. 6, property 3 listed in Sect.~\ref{s:model}). In terms of our model, we can make use of the fact the $PF$ stays almost constant down to the X-ray luminosity $L_{PF}\sim 6\times 10^{35}$~erg~s$^{-1}$  (Fig.~6). By identifying this luminosity with the critical $L_\dag$ corresponding to $\tau_v\sim 1$  for scattering, we can evaluate the required NS magnetic field $\mu_{30}\sim (3\times 10^{35} \mathrm{erg\,s}^{-1}/ L_{PF})^{10/3}\approx 0.1$. 
This estimate supports our conjecture that the NS magnetic field in \src\  is indeed lower than the canonical NS value $10^{12}$~G. We note also that the expected NS equilibrium period at the settling quasi-spherical accretion for this low NS magnetic field could be exactly in the right range of a few 100 s for the binary period 165 days, the typical X-ray luminosities and stellar wind velocities\footnote{An attempt to estimate the NS magnetic field using the NS equilibrium spin period only, $P_{eq}\sim \mu^{12/11}v_w^4$, would fail due to unknown value of the stellar wind velocity.}. 

We conclude that all significant properties 1-4 of the gross X-ray pulse profile evolution of \src\  revealed by the X-ray spectroscopy of the \nustar\ observations, might be explained coherently by the scattering of O-photons, which are mostly produced near the NS surface at low accretion rates, in the accreting plasma above one NS polar cap. However, in this case, a detailed 
pulse profile change with X-ray luminosity shown in Fig. 5 remains unexplained.

\subsubsection{Photon scatterings above two NS magnetic poles}
\label{s:2poles}

The X-ray pulse evolution of \src\ is most clearly visible in Fig. 5. The most striking feature is that the hard 12-78 keV pulse shape 
changes rather little with X-ray luminosity (the middle panel) compared to the 3-12 keV pulse (the upper panel). The latter reveals a more complicated behavior, from a sine-like shape at low luminosities transforming through a two-hump form at mediate fluxes to an almost single-pulse narrower shape at the highest luminosities. Here the left peak at the phase $\sim 0.3$ almost disappears to leave only the right peak at the phase $\sim 1.5-1.6$. 

This picture may suggest that in fact we are observing two peaks from \textit{two NS magnetic poles} (call them south (S) and north (N) poles corresponding to the left and right peaks, respectively), which is generally more likely. In the hard 12-78 keV range these two peaks almost merge to form the apparently one sine-like peak, although traces of two S and N peaks could be still discerned. 

The suppression of the low-energy S peak with increasing X-ray luminosity suggests that the 3-12 keV emission gets scattered in the accreting matter, as described above, which is possible 
if the O-X photon transformation occurs below 12 keV. This is possible if the magnetic field of the S pole is about $10^{12}$~G corresponding to CRFS at $E_{cyc}^S\sim 12$~keV in the spectrum. The scattering of X-photons from the S-beam below $E_{cyc}$ explains the enhanced hardness ratio  at the phase 0.3-0.4 at the highest luminosities (Fig. 9, the bottom right panel).

As for the N-beam that remains visible and non-scattered at all luminosities (the upper panel of Fig. 5), it may be due to a \textit{lower} magnetic field of the N pole such that $E_{cyc}^N<3$~keV so that all generated O-photons are above the CRSF energy. As no strong transformation of the N-pole has occurred up to luminosities $\sim 10^{36}$ erg s$^{-1}$, we can apply the same model as in the previous subsection to identify the critical luminosity $L_\dag$, corresponding to the unit optical depth, to estimate the NS magnetic field at this pole to be a few times $10^{10}$~G.

The different magnetic field at NS poles is not unusual for X-ray pulsars due to the possible non-dipole structure of the NS magnetic field  (see the discussion of Her X-1 in \citealt{1991SvAL...17..339S,2013MNRAS.435.1147P}). If this is true for \src\ , it can indeed be difficult to find the CRSF signature in the X-ray spectrum because at high luminosities the O-photons from the N pole with lower magnetic field dominate.

To verify this model, search for the putative 12-17 keV CRSF should be made only at low luminosities, which is challenging. Here the future X-ray polarization observations may be crucial.

\section{Conclusions}

The \nustar\ observation of the SFXT pulsar \src, performed at the expected peak of the June 2017 outburst, 
showed an ample range of variability of the flaring activity, reaching two orders of magnitude in the energy range 3-78 keV. 
This observation  has allowed us to perform the most sensitive search of \cyclabs\ in the hard X-ray spectrum of \src, to date, with null results. 
We were not able to confirm the hint of a 17 keV \cyclab\ obtained during a \nustar\ observation of the 2016 
outburst,  at a similar X-ray flux \citep{Sidoli2017}. Thus, the issue of the direct measurement of the NS magnetic field in this SFXT pulsar remains open. However, 
indirect inference about the possible value of the NS magnetic field in this source can be made from the analysis of the spectroscopic properties of X-ray pulses and their change with X-ray luminosity (see Section \ref{s:model}).

The pulsar spin period displayed no significant variability with respect to any previous measurements, with 
a profile evolving with both the energy and  the luminosity.
The pulsed fraction increased with the energy and decreased with the source X-ray intensity, within the range of luminosities probed during this observation (from an average 
L$_{\rm 1-100\,keV}$=3.5$\times10^{35}$~erg~s$^{-1}$ 
to L$_{\rm 1-100\,keV}$=2.4$\times10^{36}$~erg~s$^{-1}$).

The long spin period and  the low X-ray luminosity in outburst led us to discuss the source properties within the framework of the
quasi-spherical settling accretion model  as the physical mechanism driving the SFXT phenomenology (see \cite{Shakura2014} for more detail).
The evolution of the pulse profile at different X-ray luminosities is a further observational fact  supporting this  model. Indeed, the observed 187~s NS spin period in \src, perceived as the equilibrium spin period at the quasi-spherical settling accretion, implies a rather moderate NS magnetic field. This conjecture is simultaneously supported by an insignificant pulse phase shift, the constant hardness ratio of the X-ray light curve at different luminosities, the increase in the pulse fraction at low luminosity and the null detection of CRSF. These properties can be naturally explained by scattering of ordinary X-ray photons  in the accreting matter above NS polar caps. The estimated surface NS magnetic field in this model is about $10^{11}$~G. 

Alternatively (see Section \ref{s:2poles}), the detailed pulse profile evolution can be understood in terms of accretion onto NS magnetic poles with the surface magnetic field strength differing by one order of magnitude (e.g., due to a non-dipoles magnetic field structure). In this case, the strongest NS magnetic field was estimated to be around the standard NS field value $\sim 10^{12}$~G. The null detection CRSF in our observations at the brightest state may be due to the most of photons generated at this state being ordinary photons above the cyclotron frequency from the pole with lower magnetic field. 

The model can be further tested by searches for CRSF at low 
energies and by future X-ray polarization observations.

\begin{acknowledgements}

This work is based on data from observations with \nustar\ and \sw.
The \nustar\ mission is a 
project led by the California Institute of Technology,
managed by the Jet Propulsion Laboratory, and funded
by the National Aeronautics and Space Administration. 
This research made use of the {\em NuSTAR DAS} software
package, jointly developed by the ASDC (Italy) and Caltech (USA). 
We thank the Neil Gehrels \sw\ team and the PI for making the \sw\ monitoring observations possible.
Italian authors acknowledge the financial contribution from the agreement ASI-INAF NuSTAR I/037/12/0. Work of PK is partially supported by RFBR grant 19-02-00790.

\end{acknowledgements}

\end{document}